\documentclass[aps,prl,twocolumn,showpacs,amsmath,amssymb,floatfix,superscriptaddress,notitlepage]{revtex4-1}
\usepackage{color}
\usepackage{bbm}
\usepackage{graphicx}% Include figure files
\usepackage{dcolumn}% Align table columns on decimal point
\usepackage[utf8]{inputenc}
\usepackage{graphicx}
\usepackage{epstopdf}
\usepackage{amsthm}
\usepackage{amsmath}
\usepackage{empheq}
\usepackage{bbm}
\usepackage{braket}
\usepackage{amssymb}
\usepackage[usenames,dvipsnames]{xcolor}
\usepackage{cases}
\usepackage{latexsym}
\usepackage[colorlinks=true,citecolor=Cerulean,linkcolor=RubineRed,urlcolor=Cerulean]{hyperref}

\graphicspath{ {images/} }

\newcommand{\id}{\mathbbm{1}} %Identity
\newcommand{\tr}[1]{\operatorname{\textnormal{Tr}}\left( {#1} \right)} 
\newcommand{\trs}[2]{\operatorname{\textnormal{Tr}}_{{#1}}\left( {#2} \right)}  %Trace
\newcommand{\info}{\mathcal{I}_F}

\newcommand{\sys}{\mathcal{S}} %system
\newcommand{\bat}{\mathcal{W}} %battery
\newcommand{\bath}{\mathcal{B}} %bath
\newcommand{\anc}{\mathcal{A}} %ancillas

\newcommand{\work}{\mathcal{F}} %system
\newcommand{\LP}[1]{{\color{black}  #1}}

\newcommand{\LPthree}[1]{{\color{black}  #1}}

\begin{document}

\title{Fluctuations in extractable work bound the charging power of quantum batteries}

\author{Luis Pedro Garc\'ia-Pintos}
\thanks{Corresponding author}
\email{lpgp@umd.edu}
\affiliation{Joint Center for Quantum Information and Computer Science and Joint Quantum Institute, NIST/University of Maryland, College Park, Maryland 20742, USA}
\affiliation{Department of Physics, University of Massachusetts, Boston, Massachusetts 02125, USA}

\author{Alioscia Hamma}
\email{alioscia.hamma@umb.edu}
\affiliation{Department of Physics, University of Massachusetts, Boston, Massachusetts 02125, USA}

\author{Adolfo del Campo}
\email{adolfo.delcampo@dipc.org}
\affiliation{Donostia International Physics Center,  E-20018 San Sebasti\'an, Spain}
\affiliation{IKERBASQUE, Basque Foundation for Science, E-48013 Bilbao, Spain}
\affiliation{Department of Physics, University of Massachusetts, Boston, Massachusetts 02125, USA}

\date{\today}

\begin{abstract}

We study the connection between the charging power of quantum batteries and the fluctuations of the \LP{extractable} work. 
We prove that in order to have a non-zero rate of change of the extractable work, 
the state $\rho_\mathcal{W}$ of the battery cannot be an eigenstate of a `\emph{\LP{free energy} operator}', defined by $\mathcal{F}~\equiv~H_\mathcal{W}~+~\beta^{-1}\log(\rho_\mathcal{W})$, where $H_\mathcal{W}$ is the Hamiltonian of the battery and $\beta$ is the inverse temperature of a reference thermal bath with respect to which the extractable work is calculated. 
We do so by proving that fluctuations in the \LP{free energy operator}
%, which characterize  extractable work,}
 upper bound the charging power of a quantum battery.
 Our findings also suggest that quantum coherence in the battery enhances the charging process, which we illustrate on a toy model of a heat engine.
 
 \
 
 \noindent [\emph{Note: this version includes our Reply to a Comment by Cusumano and Rudnicki, both published in Phys. Rev. Lett.}]

\end{abstract}

\maketitle

The study of the thermodynamics of quantum systems can be traced back to the beginnings of quantum theory, but a recent surge in interest has been spurred by major advances~\cite{anders2017focus} 
in which the use of quantum information theoretic tools  plays a key role \cite{Allahverdyan04,workextractionBrandaoPRL13,
workextractionAbergPRL14,Huberreviewthermo2016,
MasanesNatComm2016work,masanes2017general,2019arXiv190202357A}.
Despite this progress, fundamental issues continue to be a matter of debate. Prominent examples include  the notion of  thermodynamical work in the quantum 
realm~\cite{aaberg2013truly,FrenzelPRE2014,
jarzynski2015quantum,AlhambraPRX2016,
gallego2016thermodynamic}, 
and the role of quantum coherences and their potential use as a thermodynamical
 resource~\cite{brandner2015coherence,lostaglio2015quantum,
 UzdinPRX2015,korzekwa2016extraction,
 CoherenceRevMod2017,FrancicaPRE2019,
PetruccioneSciRep2019,PoliniPRL2019}. 

In this context, the study of heat engines and quantum machines has proven useful, helping to illustrate the capabilities and limits of quantum devices ~\cite{BrunnerPRE2012,LindenPRL2010,CorreaPRE2013,
PopescuNatComm2014,Kosloff2014,
AdC2014,
GelbwaserNJP2015,BrunnerPRB2016,
GelbwaserNJP2016,caravelli2019random}.
The analysis of these devices has established  a trade-off between work fluctuations and dissipation in heat engines~\cite{Masahito2015,Funo17}, an increase in the charging power~\cite{Binder15,AdCNJP2016,
ModiPRL2017,FerraroPRL2018}
and decrease in fluctuations~\cite{Perarnau_Llobet_2019}
with collective operations acting on
storage devices in parallel, the advantage on charging power over many cycles of a heat engine~\cite{AdCPRL2017cycleengine} and of indistinguishable heat engines \cite{Watanabe19}, as well as an increased efficiency when considering correlated thermal machines~\cite{MuellerCorrelatedMachines2017}.
On the other hand, it has also been shown that entanglement is not indispensable for work extraction~\cite{KarenPRL2013}, and that within the set of Gaussian operations to charge a battery there are inevitably fluctuations in the stored work~\cite{HuberGaussianBatteries2017}.

Along parallel lines, Mandalstam and Tamm considered the limits that quantum mechanics imposes on the rate of quantum evolution~\cite{mandelstamtamm1945}.
By considering the minimum time necessary for a state to evolve to an orthogonal one, their work inspired a wide range of results~\cite{YakirPRL1990,
margoluslevitin1998,LloydNature2000,LloydPRL2002,
MacconePRA2003,DavidovichPRL2013,delCampoPRL2013,
DeffnerLutzPRL2013,MarvianPRA2016,Shanahan18},
that are usually encompassed under the term of quantum speed limits to evolution \cite{DeffnerCampbell17}.  
%In particular, limits to the speed of evolution have been connected to the presence of quantum coherence~\cite{MarvianPRA2016},  shown to 
%impose constraints on parameter estimation~\cite{demkowicz2012NatComm,BeauPRL2017} and quantum control~\cite{GiovanettiPRL2009,
%AdCPRL2012}, and provide ultimate limits on the processing power of physical systems~\cite{margoluslevitin1998,LloydNature2000,LloydPRL2002,
%MacconePRA2003}. 

In quantum thermodynamics, speed limits were first discussed to bound the output power of heat engines \cite{AdC2014}. They are known to govern work and energy fluctuations in superadiabatic processes \cite{An16,Funo17,DeffnerPRL2017} and can be used to analyze advantages from many-particle effects in quantum batteries \cite{Binder15,ModiPRL2017}. 
Articles studying the limits to increasing the energy of batteries in unitary protocols  appeared recently~\cite{Ito2017arXiv,PoliniPRB2018,
LewensteinBatteries18,PoliniPRB2019}. 
In~\cite{MasanesNatComm2016work}, general bounds on the efficiency of an engine and the extractable work are derived as a function of the work fluctuations, for a battery modeled by a weight that starts and ends in classical (diagonal) states.

In this Letter,
we investigate the fundamental limits that quantum mechanics imposes on the process of charging a battery, \LPthree{where we take a battery to be any system that can store energy for subsequent extraction. We}
%%%%%%%%% and
 show that in order to have a non-zero charging power there must exist fluctuations
of a free energy operator that characterizes the extractable work in the battery.  
%In this way charging power and statistics 
% in the extractable work that is stored in the battery. 
In particular, we provide bounds on the charging power that a) explicitly take into consideration the entropic aspects of the amount of \LP{extractable work} as opposed to limiting the study to energy storage, and b) are valid for arbitrary states and a wide range of protocols, including those based on unitary evolution as well as protocols involving open-system dynamics, and hold regardless of the nature of the battery, its state, and the properties of the \LP{extractable work of the battery}.
 
In a work extraction protocol any system out of thermal equilibrium is a resource, from which work can be extracted if one has access to a thermal bath.
If the resource system is in state $\rho$, and $\tau_\beta$ denotes the thermal state at inverse temperature $\beta$,
% the \LP{maximum amount of work that can be extracted, on average,} from asymptotically many copies of $\rho$ is given by
 the \LP{maximum amount of work that can be extracted} from %asymptotically many copies of 
 the resource state $\rho$ is given by
\begin{align}
\label{eq:workfreenergy}
W_\textnormal{max}  = F(\rho) - F(\tau_\beta),
\end{align}
where the free energy is $F(\rho) = U - S/\beta$, and $U$ and $S = -\tr{\rho \ln\rho}$ are the \LP{average} energy and von Neumann entropy of the system, respectively~\cite{workextractionBrandaoPRL13,
workextractionAbergPRL14,
workextractionPopescuNatComm14,
MullerPRX2018}.
We refer to $W_\textnormal{max}$ as the `\emph{extractable work}'. 
In~\cite{workextractionBrandaoPRL13} this is shown to be the case, on average, in the limit of asymptotically many copies of $\rho$, while~\cite{workextractionPopescuNatComm14} constructs a protocol that works for a single copy of $\rho$, albeit with non-zero fluctuations in the extracted work and with a toy-model as work reservoir. Interestingly,~\cite{MullerPRX2018} proves that, if the resource system is allowed to become correlated to catalytic systems, then $W_\textnormal{max}$ exactly determines the extractable work in single-shot scenarios as well. 
It is worth clarifying that the notion of work used in~\cite{workextractionBrandaoPRL13,
workextractionAbergPRL14,
workextractionPopescuNatComm14,
MullerPRX2018} is an operational one, in which the extracted work is stored in reservoirs, for posterior extraction.
 Specifically,~\cite{workextractionBrandaoPRL13,MullerPRX2018} show that an amount of energy equal to the extractable work can be used to excite two-level systems from the ground state, while in~\cite{workextractionAbergPRL14,
 workextractionPopescuNatComm14} it is used to raise the energy of an (unbounded) harmonic oscillator. 
Importantly, these protocols also provide a mechanism to extract the same amount of energy from the reservoirs, 
%. Thus, this 
thus providing an operational way in which work is defined in this optimal-protocol context 
\LPthree{(albeit with the possible need of auxiliary resources, such as a system that serves as entropy sink in~\cite{MullerPRX2018}).}
%, slowly growing xxx in [REF], and xxxx in []). xxx}
%%%%%This work can then be used to drive another physical process, say lifting a weight or lighting a bulb. 
%%%%%However, in most practical situations one wants to store the extracted work for later consumption.

\begin{figure}[t]
  \centering  
        \includegraphics[trim=00 00 00 00,width=0.4
        \textwidth]{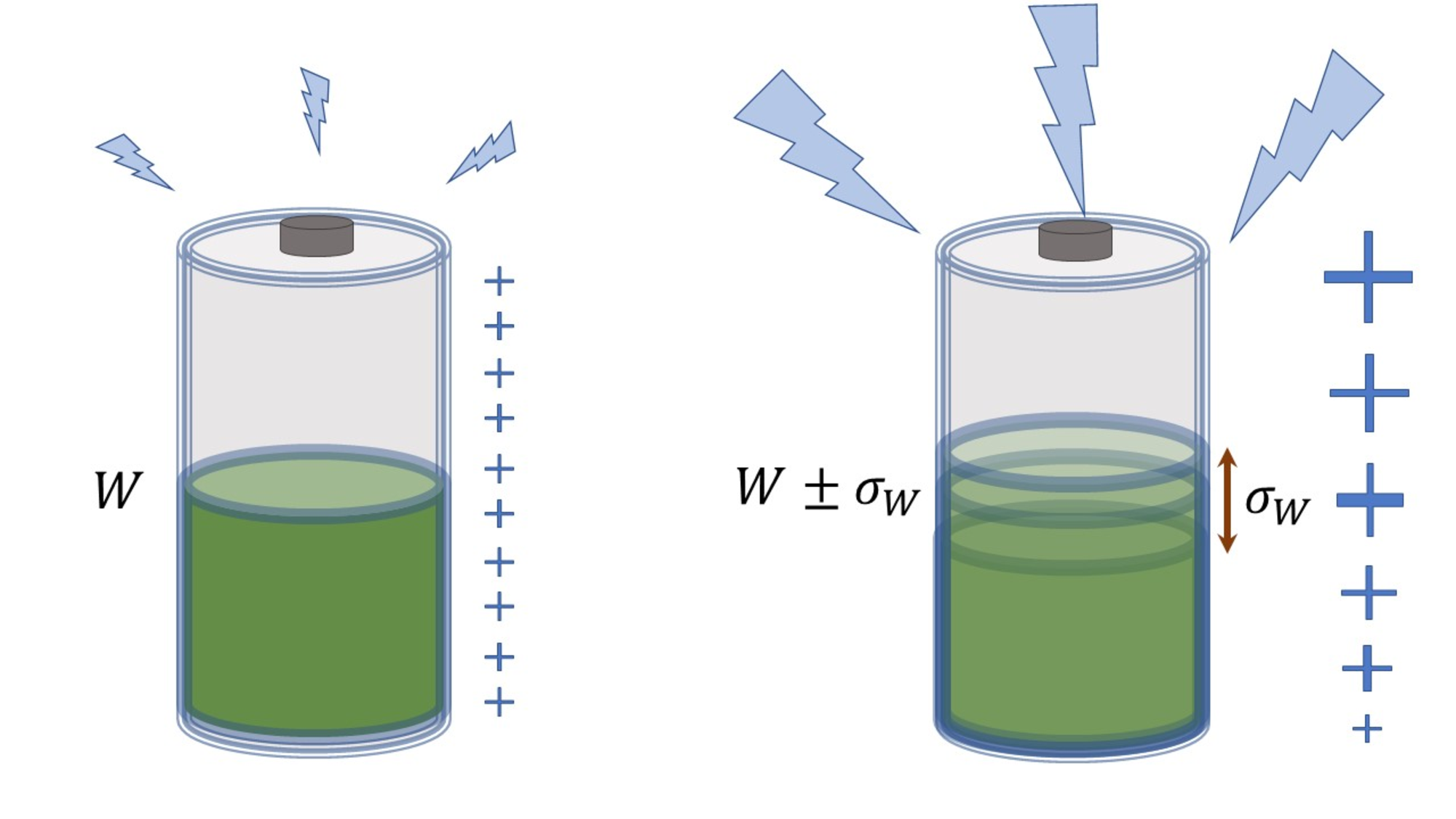} 
\caption{\label{fig:artwork}
\textbf{ Fluctuations in extractable work  limit the charging power of batteries.}
A quantum battery is a system that follows the laws of quantum physics, and that can store energy  to be extracted later in order to perform work.
The most important quantity in a  battery is the  maximum amount of work $W$ that can be extracted from it, i.e., the extractable work, which in certain cases is directly characterized by the free energy.
However, due to the quantum nature of the battery, this free energy is a random quantity with uncertainties. 
We prove that the rate at which any battery can be charged is linked to such fluctuations, via fundamental bounds that hold for any charging protocol.
 For a battery with a well-defined value of the  extractable work $W$, depicted on the left, the maximum rate at which the extractable work changes
  is zero. 
  Fluctuations allow for faster charging rate, as illustrated on the right. %A battery 
}
\end{figure}

In this Letter we focus on the  limitations that quantum mechanics imposes on any  process of storing or extracting the extractable work, as illustrated in Fig.~\ref{fig:artwork}.
 
\

\noindent \emph{Isolated system analysis.---} 
A protocol to extract work from a system and store it in a battery necessarily couples, at some point, the battery with the bath, the system, and/or auxiliary ancillas $\anc$ used in order to implement the necessary operations. For instance,  ancillas can be used to describe an arbitrary map acting on the system. 
We denote the system used \LP{as work reservoir}, i.e. the \emph{battery}, by $\bat$, and
its Hamiltonian by $H_\bat$.
We assume that the total system, consisting of all of the above, is isolated and evolves unitarily with a total Hamiltonian
\begin{align}
\label{eq:hamiltonian}
H & = H_{\sys\bath\anc}\otimes \id_\bat + \id_{\sys \bath \anc} \otimes H_\bat + V ,
\end{align}
where $H_{\sys\bath\anc}$ accounts for all terms that correspond to bath, system and ancilla evolution, $V $ incorporates all terms corresponding to interactions with the battery, and $\id_{\{\cdot\}}$ denotes identities on the subspaces indicated by the subindex. 
Note that, while we allow $H_{\sys \bath \anc}$ and $V $ to be explicitly time-dependent, we assume $H_\bat$ to be time-independent, motivated by the idea of having a battery whose physical characteristics --e.g. Hamiltonian-- do not change.
Then, the full state $\rho$ of $\sys \bath \anc \bat$ evolves according to
\begin{align}
\label{eq:isolatedevol}
\frac{d\rho }{dt} = -i [H,\rho ].
\end{align} 
For ease of notation we omit the explicit time-dependence of the evolving state $\rho$. 

A central quantity behind the results of this Letter is the `\emph{free energy operator}', defined with respect to a bath at inverse temperature $\beta$ as 
\begin{align}
\mathcal{F} \equiv H_\bat + \beta^{-1} \log \rho_\bat,
\end{align}
where $\rho_\bat = \trs{\sys \bath \anc}{\rho}$ is the state of the battery.
The free energy operator $\work$ characterizes the amount of work that can be extracted from the battery, as from Eq.~\eqref{eq:workfreenergy} it follows that
\begin{align}
W_\textnormal{max}  = \langle \mathcal{F} \rangle_\bat - \langle \mathcal{F} \rangle_\beta,
\end{align}
where $\langle \mathcal{F} \rangle_\bat \equiv \tr{\mathcal{F} \rho_\bat}$ and $\langle \mathcal{F} \rangle_\beta \equiv \tr{\mathcal{F} \tau_\beta}$. 
 Note that, while $\mathcal{F}$ is a Hermitian operator, it depends on the state of the battery and it need not correspond to a physical observable.
It is also worth stressing, though, that $\work$ is distinctively different to the process-dependent `work operator' considered in~\cite{AllahverdyanWORKOPPRE2005,
TalknerWORKOPPRE2007,MartiWORKPRL2017}. Unlike the `work operator', $\work$ corresponds to a state function, and its average
 characterizes the \emph{extractable work}~\cite{workextractionBrandaoPRL13,workextractionAbergPRL14,
workextractionPopescuNatComm14,
MullerPRX2018}.

The rate at which the battery's extractable work changes, that we refer to as the \emph{charging power}, is given by
\begin{align}
P(t) &=\frac{d W_\textnormal{max}  }{ dt} =\frac{d \langle \mathcal{F} \rangle_\bat }{dt} \\ &= \frac{d \tr{\rho_\bat  H_\bat}}{dt} - \frac{1}{\beta} \frac{d S(\rho_\bat)}{ dt },
\end{align}
where we used that $H_\bat$ is time-independent.
 As proven in~\cite{Aspuru-GuzikPRL2011,DasJMathPhys2018}, it holds that $ d S(\rho_\bat)/dt=-i\tr{V\left[ \log\rho_\bat \otimes \id_{\sys \bath \anc}, \rho \right]}$. Note that $P(t)$ limits the power needed in realistic charging protocols. Specifically, the energy $E$ needed to increase a battery's extractable work by  $\Delta W  \equiv W_\textnormal{max}(t) - W_\textnormal{max}(0)$ 
 in realistic charging protocols is bounded by the charging power, as  $E \geq \Delta W = \int_0^t P(t') dt'$.

Using Eqs.~\eqref{eq:hamiltonian} and~\eqref{eq:isolatedevol},
it further follows that
\LPthree{
$d \tr{\rho_\bat  H_\bat}/dt =-i \tr{[\rho, H_\bat\otimes \id_{\sys \bath \anc}]V}$}. 
As a result,
\begin{align}
P(t) =& -i \tr{[\rho , H_\bat \otimes \id_{\sys \bath \anc} ] V } \nonumber \\
&- i \frac{1}{\beta}\tr{\left[ \rho , \log\rho_\bat\otimes \id_{\sys \bath \anc} \right]V}  \nonumber 
\\ =& -i \tr{\left[ \rho , \mathcal{F} \otimes \id_{\sys \bath \anc} \right] V} \label{eq:coherenceBattery}.
%\\ & = i \tr{\left[ \rho , V \right] %\mathcal{F} \otimes \id_{\sys \bath \anc}} %\label{eq:coherenceInteraction}.
\end{align}
At a qualitative level, this simple derivation 
suggests that coherence in the eigenbasis of the \LP{free energy operator --which in turn necessitates coherence in the energy eigenbasis--} serves to enhance the charging process.
 Indeed, if one considers a battery initially uncorrelated from other systems, $\rho = \rho_\bat \otimes \rho_{\sys \bath \anc}$, the charging power is zero unless  the state is coherent in the eigenbasis of $\mathcal{F}$.

Our main result sets bounds on the charging power. Defining $\delta \mathcal{F} \equiv \mathcal{F} -\langle \mathcal{F} \rangle_\bat$ and $\delta V = V - \langle V \rangle$, it holds that
\begin{align}
\label{eq:powerboundderivation}
P^2(t) &\le \Big| \tr{\rho  \left[ \delta\work \otimes \id_{\sys \bath \anc} , \delta V \right]} \Big|^2 \le 4 \Big| \tr{\rho  \, \delta\work \,  \delta V    } \Big|^2 \nonumber \\
&\le 4 \tr{ \rho  \left(\delta\work \right)^2 \otimes \id_{\sys \bath \anc} } \tr{ \rho  \left( \delta V  \right)^2 },
\end{align}
where  we use that $H_\bat$ is time-independent and denote the averages of the extractable work and battery interaction energy by  $\langle \work \rangle_\bat = \tr{\rho_\bat  \work}$  and $\langle V \rangle  \equiv \tr{\rho  V }$, respectively.
For the last step of the calculation we used the fact that for hermitian operators $A > 0$, $B$ and $C$, the Cauchy-Schwarz inequality implies that $\text{Tr}^2\left(ABC\right) =\text{Tr}^2\left(\sqrt{A}BC\sqrt{A}\right)\leq   \tr{AB^2}\tr{AC^2}$.
 Denoting the standard deviations of $\work$ and of the battery interaction Hamiltonian by $\sigma_\work$ and $\sigma_V$  respectively, 
\begin{align}
\sigma_\work^2(t) &\equiv \left\langle \work^2 \right\rangle_\bat - \left\langle \work \right\rangle_\bat^2  \\
\sigma_V^2(t) &\equiv \left\langle V^2 \right\rangle - \left\langle V \right\rangle^2,
\end{align}
Eq.~\eqref{eq:powerboundderivation} implies
\begin{align}
\label{eq:powerbound}
|P(t)| \le 2 \sigma_\work(t) \sigma_V(t).
\end{align}

This proves the existence of a trade-off between charging power and the fluctuations of the  free energy operator $\work$:
for a fixed interaction with the battery, a desired power input necessarily comes with fluctuations of the  operator $\work$ whose mean characterizes the extractable work of the battery.  
In contrast, attempting to charge a battery with a deterministic amount of \LP{extractable work}, such that $\sigma_\work = 0$, leads to a null instantaneous charging power. 
\LP{It is straightforward to see that this implies, for instance, zero charging power for batteries in eigenstates of energy.}
We stress that this bound applies, in particular, to protocols that charge the battery via unitary evolution with a time-dependent perturbation Hamiltonian $V(t)$~\cite{Binder15,AdCNJP2016,
ModiPRL2017,Ito2017arXiv,PoliniPRB2018,
LewensteinBatteries18,
caravelli2019random}.

\

\noindent \emph{Open system analysis.---}
Evaluating bound~\eqref{eq:powerbound}
for charging protocols involving unitary, time-dependent, control of the battery is straightforward, as it solely involves knowledge of the state of the battery, its Hamiltonian $H_\bat$, and the control Hamiltonian $V$. 
However, 
evaluating the factor $\sigma_V(t)$ 
may be hard
for protocols that involve contact of the battery with secondary systems, as evaluating it requires specifying the full state $\rho$ of the battery and all the systems it interacts with. 
In the light of this, it is of practical importance to extend the analysis to encompass open-system descriptions of the dynamics of the battery, which we do next.

The formalism introduced so far allows to extend the analysis to include an open-system description of the dynamics of the battery. 
From Eq.~\eqref{eq:isolatedevol}, the state of the battery evolves according to 
\begin{align}
\label{eq:openevol}
\frac{d}{dt} \rho_\bat &= -i \trs{\sys \bath \anc}{\left [H,\rho  \right]} \nonumber \\
%&= -i \trs{\sys \bath \anc}{\left [ \id_{\sys \bath \anc} \otimes H_\bat + V,\rho  \right]} \nonumber \\
& = -i[H_\bat,\rho_\bat ] -i \trs{\sys \bath \anc}{\left [ V,\rho  \right]}.
\end{align}
In the Markovian and weak-coupling limits the evolution of a system interacting with an environment is well approximated by 
\begin{align}
\frac{d}{dt} \rho_\bat  &\approx -i[H_\bat,\rho_\bat ] -i[\tilde H_\bat,\rho_\bat ]  \\
&+ \sum_j \gamma_j \left( L_j \rho_\bat  L_j^\dag - \frac{1}{2} \left\{ L_j^\dag L_j ,\rho_\bat  \right\} \right) ,\nonumber
\end{align}
where $\tilde H_\bat$ accounts for the unitary part of the evolution due to the interactions, 
the Lindblad operators $L_j$ characterize the non-unitary effect of the interaction of the battery with the remaining systems, and
the rates $\gamma_j$ are non-negative~\cite{Book-Open}.
 Then,
\begin{align}
& \trs{\sys \bath \anc}{\left [ V,\rho  \right]} \approx [\tilde H_\bat,\rho_\bat ]  \\
 &\quad \quad \quad  + i \sum_j \gamma_j \left( L_j^\dag \rho_\bat  L_j - \frac{1}{2} \left\{ L_j^\dag L_j ,\rho_\bat  \right\} \right). \nonumber
\end{align}

With this, we prove in the Supplemental Material~\cite{SM} that the rate at which \LP{the extractable work of the battery changes} is upper bounded by
\begin{align}
\label{eq:boundopen}
| P(t) | &\leq  2 \sigma_{\work}(t) \sigma_{\tilde H_\bat}(t) + \sum_j   \gamma_j  \sqrt{\left\langle \big| [ \delta \work, L_j ] \big|^2 \right\rangle} \|L_j \|,
\end{align}
\LP{where the operator norm $\| A \|$
 is given by the largest modulus of the eigenvalues of an operator $A$}, and $|A|^2 = AA^\dag$. 
This sets a bound 
valid for charging protocols based on open-system approaches. Importantly, the bound depends solely on the state $\rho_\bat$ of the battery, and the decay rates $\gamma_j$ and operators $\widetilde{H}$ and $L_j$, fixed by the master equation that governs the dynamics of the battery.

Remarkably, for the open-system case it also holds that the charging power is null unless there exist fluctuations in the \LP{free energy operator $\work$.}  
  In order to see this, let $\ket{j}$ denote the eigenbasis of $\delta \work$, with $\delta \work = \sum_j w_j \ket{j} \bra{j}$,  $\rho_{jk} = \bra{j} \rho_\bat \ket{k}$ and $L_{jk} = \bra{j} L \ket{k}$. One then finds that
\begin{align}
\left\langle \big| [ \delta \work, L_j ] \big|^2 \right\rangle = \sum_{jkl} \rho_{jk} L_{kl} L^\dag_{lj} \left( w_l^2 - w_l w_j - w_l w_k + w_j w_k \right).
\end{align}
As a result, for states $\rho_\bat = \ket{j} \bra{j}$ with a deterministic amount of \LP{free energy characterized by} 
%\LP{extractable work}
 $w_j$, one has that $\left\langle \big| [ \delta \work, L_j ] \big|^2 \right\rangle = 0$ and that $\sigma_\work = 0$, leading to a null charging rate
$|P(t)|~=~0$.
By contrast, states with support on more than one eigenstate of $\work$ can sustain a higher charging power, as both $\left\langle \big| [ \delta \work, L_j ] \big|^2 \right\rangle \neq 0$ and $\sigma_\work \neq 0$.

Finally, in the case of Hermitian Lindblad operators $L_j = L_j^\dag$ we prove in the Supplemental Material~\cite{SM} that the bound~\eqref{eq:boundopen} simplifies to 
\begin{align}
\label{eq:boundopenHermi}
| P(t) | &\leq  \sigma_{\work}(t) \left( 2 \sigma_{\tilde H_\bat}(t) + \sum_j   2 \gamma_j  \|L_j \|^2 \right).
\end{align}

\

\noindent \emph{Illustration with a heat engine.---}
We consider a minimal, self-contained, model of a heat engine that stores work in a quantum battery, as studied in detail in~\cite{Linden2010,Linden2010PRE}. 
The engine consists thermal baths at different temperatures, $T_h$ and $T_c$, as a resource. Note that these baths are internal to the workings of the protocol to extract energy; in particular, the temperatures $T_h$ and $T_c$ are unrelated to the inverse temperature $\beta$ of the reference bath with respect to which work is defined in Eq.~\eqref{eq:workfreenergy}.  

In the engine, heat flow from the hot to the cold bath is exploited to extract work and store it in \LP{a toy-model battery that} consists of a harmonic oscillator unbounded from below, with an energy gap $\epsilon$ for a Hamiltonian
\begin{align}
H_\bat = \sum_{n = -\infty}^\infty n\epsilon \ket{n}_w\!\bra{n}.
\end{align}
The storage device is indirectly coupled to the heat baths via a `switch', consisting of two qubits. Qubit $1$, with energy gap $E_1$, is coupled to the cold bath, while qubit $2$, with energy gap $E_2$, is coupled to the hot bath. 
The qubits have a free Hamiltonian given by
$H_\sys = E_1 \ket{1}_1\!\bra{1} + E_2 \ket{1}_2\!\bra{1}$,
with energies taken such that $E_2-E_1 = \epsilon$, and they interact with the battery via
\begin{align}
V = g \sum_{n } \Big( \ket{01,n}\!\bra{10,n+1} + \ket{10,n+1}\!\bra{01,n} \Big),
\end{align}
where $g$ is a coupling constant.
Qubits $1$ and $2$ are assumed to thermalize due to the interaction with the thermal baths at rates $p_1$ and $p_2$, respectively.
With these considerations, the evolution of the mean energy in the battery is solved analytically in~\cite{Linden2010,Linden2010PRE}, where it is found that
choosing the parameters of the model correctly make the device work as a heat engine, storing work in $\bat$.

We consider for simplicity a reference bath of zero temperature to calculate the \LP{extractable work of the battery}. For this case, we derive the equations governing the dynamics of the charging power, \LP{extractable} work, and its fluctuations \LP{in the Supplemental Material~\cite{SM}}.
We are interested in the charging power \LP{for initial states with uncertain amounts of free energy.}

\begin{figure}[t]
\begin{center}
\includegraphics[width=0.5051\linewidth]{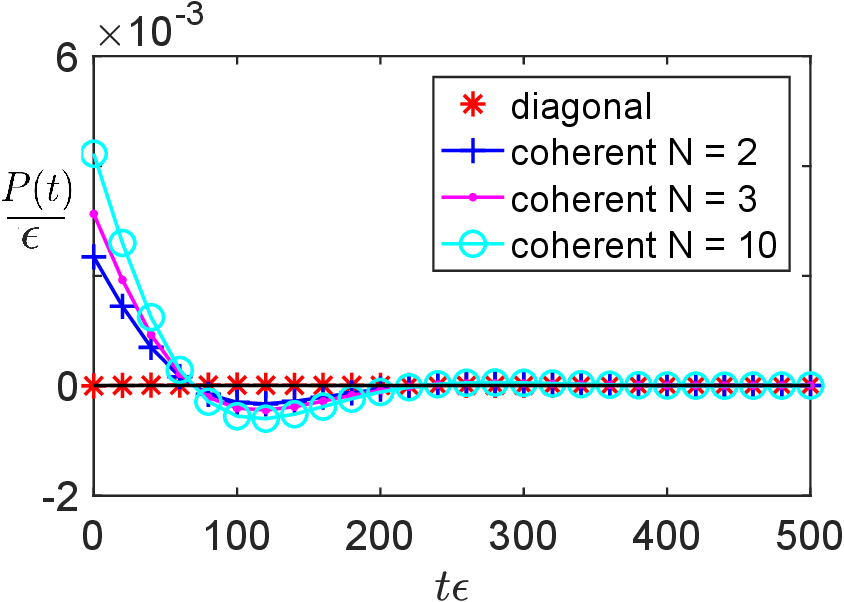}
\includegraphics[width=0.48\linewidth]{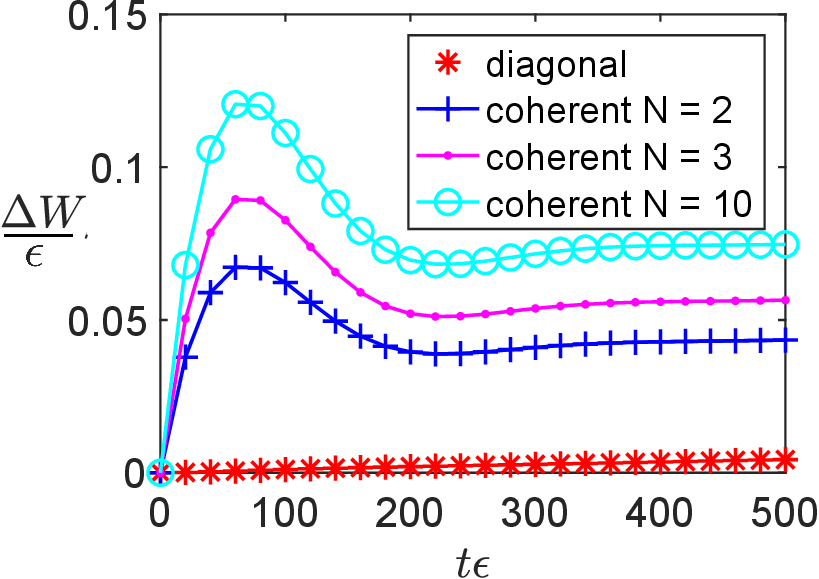}
\end{center}
\caption{\label{fig:powerandwork}
                \textbf{Charging power and \LP{extractable} work.}
Charging power $P(t)$ \LPthree{(left)} and change in  extractable work $\Delta W  \equiv W_\textnormal{max}(t) - W_\textnormal{max}(0)$ \LPthree{(right)} as a function of time for various initial states (for the values of the parameters used see~\cite{SM}). The incoherent state $\rho_\textnormal{diag}$ (red asterisk) has null charging power initially, as expected from bound~\eqref{eq:powerbound}. This charging power is slowly increased as the state supports higher fluctuations \LP{in extractable work}. A significant increase in the charging power is seen for a coherent superposition of eigenstates of $\work$. For intermediate times this initial increase is penalized with a regime in which the battery is being discharged. Nevertheless, there is a net increase in the \LP{extractable} work with respect to an incoherent initial state. Taking coherent superpositions between more than two states gives a further increase in charging power and \LP{extractable} work.
}

\end{figure}

Consider first a state without uncertainty in the $\work$.
 For instance, consider qubits $1$ and $2$ initially in states $\tau_1$ and $\tau_2$ in thermal equilibrium with their respective thermal baths, and the battery in an eigenstate $\ket{0}_\bat$ of its Hamiltonian, 
\begin{align}
\rho_{\textnormal{diag}} &= \tau_1 \otimes \tau_2 \otimes \ket{0}_\bat\bra{0}.
\end{align}
Naively, this 
 would appear to be 
an ideal initial state for the battery, with a well-defined deterministic initial amount of \LP{energy and extractable work, without fluctuations.}
However, for such a state, diagonal in both interaction and battery Hamiltonians, Eq.~
\eqref{eq:coherenceBattery}
implies that the engine initially functions with null power. This is illustrated in Fig.~\ref{fig:powerandwork}. 

In order to have non-zero charging power for product states, a coherent superposition in the \LP{free energy operator} and interaction Hamiltonian is needed.
Consequently, we consider both the qubits and battery in a pure state, the latter in a superposition between $N$ energy levels $\ket{n}_\bat$, with equal weights for simplicity:
\begin{align}
\ket{\Psi_N} &= \Big( \sqrt{r_1}\ket{0}_1 + e^{i \theta} \sqrt{1-r_1}\ket{1}_1  \Big) \nonumber \\
&\Big( \sqrt{r_2}\ket{0}_2 + \sqrt{1-r_2}\ket{1}_2  \Big) \frac{1}{\sqrt{N}} \sum_{n=0}^{N-1}\ket{n}_\bat .
\end{align}
The phase $e^{i \theta}$ fixes whether the device works as an engine or a refrigerator.

With such coherent superposition as the initial state, one has $\sigma_\work \ge 0$ and $\sigma_v \ge 0$, such that inequality~
\eqref{eq:coherenceBattery} allows to have a non-zero charging power. This is to be compared with the null charging power of state $\rho_\textnormal{diag}$. 
Figure~\ref{fig:powerandwork} compares the charging power $P(t)$ and the change in the \LP{extractable} work $\Delta W \equiv W_\textnormal{max}(t) - W_\textnormal{max}(0)$ as a function of time, for initial states given by $\rho_\textnormal{diag}$ and $\ket{\Psi_N}\!\bra{\Psi_N}$, for different values of $N$. 
A superposition between eigenstates of $\work$  causes an increase in the charging power, with respect to the null charging power of the incoherent state. 
Superpositions between more levels results in an even higher power.
This is reflected in the total \LP{extractable work as well}, with a considerable increase for coherent superpositions, with the most noticeable advantage achieved when going from incoherent state to $\ket{\Psi_2}\!\bra{\Psi_2}$.

\

\noindent \emph{Discussion.---} 
We have established a direct relationship between \LP{the statistics of the extractable work and the charging power of batteries: for the latter to be non-zero 
the extractable work of the battery has to fluctuate}. 
To this end, we have introduced bounds to the charging power in terms of \LP{a `free energy operator'} $\work~\equiv~H_\bat + \beta^{-1} \log(\rho_\bat)$ whose expectation value equals the amount of \LP{maximum amount of} work that can be extracted from the battery. Batteries in an eigenstate of $\work$ suffer from a null charging power, while fluctuations of $\work$ allow for higher charging rates.
The bounds hold for a variety of battery charging protocols, including unitary dynamics of the battery via time-dependent perturbation Hamiltonians, isolated dynamics for battery-system-bath, as well as open-system dynamics. 
Our results also identify coherence as a resource in the process of work storage and should be of relevance to the engineering of quantum thermodynamic devices in a variety of platforms.

\

This work was funded by the John Templeton Foundation, UMass Boston Project No. P20150000029279, and DOE Grant No. DE-SC0019515. 
LPGP also acknowledges partial support by AFOSR MURI project ``Scalable Certification of Quantum Computing Devices and Networks'', DoE ASCR FAR-QC (award No. DE-SC0020312), DoE BES Materials and Chemical Sciences Research for Quantum Information Science program (award No. DE-SC0019449), DoE ASCR Quantum Testbed Pathfinder program (award No. DE-SC0019040), NSF PFCQC program, AFOSR, ARO MURI, ARL CDQI, and NSF PFC at JQI.

\bibliography{referencespowerlimits}

\begin{thebibliography}{81}%
\makeatletter
\providecommand \@ifxundefined [1]{%
 \@ifx{#1\undefined}
}%
\providecommand \@ifnum [1]{%
 \ifnum #1\expandafter \@firstoftwo
 \else \expandafter \@secondoftwo
 \fi
}%
\providecommand \@ifx [1]{%
 \ifx #1\expandafter \@firstoftwo
 \else \expandafter \@secondoftwo
 \fi
}%
\providecommand \natexlab [1]{#1}%
\providecommand \enquote  [1]{``#1''}%
\providecommand \bibnamefont  [1]{#1}%
\providecommand \bibfnamefont [1]{#1}%
\providecommand \citenamefont [1]{#1}%
\providecommand \href@noop [0]{\@secondoftwo}%
\providecommand \href [0]{\begingroup \@sanitize@url \@href}%
\providecommand \@href[1]{\@@startlink{#1}\@@href}%
\providecommand \@@href[1]{\endgroup#1\@@endlink}%
\providecommand \@sanitize@url [0]{\catcode `\\12\catcode `\$12\catcode
  `\&12\catcode `\#12\catcode `\^12\catcode `\_12\catcode `\%12\relax}%
\providecommand \@@startlink[1]{}%
\providecommand \@@endlink[0]{}%
\providecommand \url  [0]{\begingroup\@sanitize@url \@url }%
\providecommand \@url [1]{\endgroup\@href {#1}{\urlprefix }}%
\providecommand \urlprefix  [0]{URL }%
\providecommand \Eprint [0]{\href }%
\providecommand \doibase [0]{http://dx.doi.org/}%
\providecommand \selectlanguage [0]{\@gobble}%
\providecommand \bibinfo  [0]{\@secondoftwo}%
\providecommand \bibfield  [0]{\@secondoftwo}%
\providecommand \translation [1]{[#1]}%
\providecommand \BibitemOpen [0]{}%
\providecommand \bibitemStop [0]{}%
\providecommand \bibitemNoStop [0]{.\EOS\space}%
\providecommand \EOS [0]{\spacefactor3000\relax}%
\providecommand \BibitemShut  [1]{\csname bibitem#1\endcsname}%
\let\auto@bib@innerbib\@empty
%</preamble>
\bibitem [{\citenamefont {Anders}\ and\ \citenamefont
  {Esposito}(2017)}]{anders2017focus}%
  \BibitemOpen
  \bibfield  {author} {\bibinfo {author} {\bibfnamefont {J.}~\bibnamefont
  {Anders}}\ and\ \bibinfo {author} {\bibfnamefont {M.}~\bibnamefont
  {Esposito}},\ }\href {\doibase 10.1088/1367-2630/19/1/010201} {\bibfield
  {journal} {\bibinfo  {journal} {New Journal of Physics}\ }\textbf {\bibinfo
  {volume} {19}},\ \bibinfo {pages} {010201} (\bibinfo {year}
  {2017})}\BibitemShut {NoStop}%
\bibitem [{\citenamefont {Allahverdyan}\ \emph {et~al.}(2004)\citenamefont
  {Allahverdyan}, \citenamefont {Balian},\ and\ \citenamefont
  {Nieuwenhuizen}}]{Allahverdyan04}%
  \BibitemOpen
  \bibfield  {author} {\bibinfo {author} {\bibfnamefont {A.~E.}\ \bibnamefont
  {Allahverdyan}}, \bibinfo {author} {\bibfnamefont {R.}~\bibnamefont
  {Balian}}, \ and\ \bibinfo {author} {\bibfnamefont {T.~M.}\ \bibnamefont
  {Nieuwenhuizen}},\ }\href {\doibase 10.1209/epl/i2004-10101-2} {\bibfield
  {journal} {\bibinfo  {journal} {Europhysics Letters ({EPL})}\ }\textbf
  {\bibinfo {volume} {67}},\ \bibinfo {pages} {565} (\bibinfo {year}
  {2004})}\BibitemShut {NoStop}%
\bibitem [{\citenamefont {Brand\~ao}\ \emph {et~al.}(2013)\citenamefont
  {Brand\~ao}, \citenamefont {Horodecki}, \citenamefont {Oppenheim},
  \citenamefont {Renes},\ and\ \citenamefont
  {Spekkens}}]{workextractionBrandaoPRL13}%
  \BibitemOpen
  \bibfield  {author} {\bibinfo {author} {\bibfnamefont {F.~G. S.~L.}\
  \bibnamefont {Brand\~ao}}, \bibinfo {author} {\bibfnamefont {M.}~\bibnamefont
  {Horodecki}}, \bibinfo {author} {\bibfnamefont {J.}~\bibnamefont
  {Oppenheim}}, \bibinfo {author} {\bibfnamefont {J.~M.}\ \bibnamefont
  {Renes}}, \ and\ \bibinfo {author} {\bibfnamefont {R.~W.}\ \bibnamefont
  {Spekkens}},\ }\href {\doibase 10.1103/PhysRevLett.111.250404} {\bibfield
  {journal} {\bibinfo  {journal} {Phys. Rev. Lett.}\ }\textbf {\bibinfo
  {volume} {111}},\ \bibinfo {pages} {250404} (\bibinfo {year}
  {2013})}\BibitemShut {NoStop}%
\bibitem [{\citenamefont {\AA{}berg}(2014)}]{workextractionAbergPRL14}%
  \BibitemOpen
  \bibfield  {author} {\bibinfo {author} {\bibfnamefont {J.}~\bibnamefont
  {\AA{}berg}},\ }\href {\doibase 10.1103/PhysRevLett.113.150402} {\bibfield
  {journal} {\bibinfo  {journal} {Phys. Rev. Lett.}\ }\textbf {\bibinfo
  {volume} {113}},\ \bibinfo {pages} {150402} (\bibinfo {year}
  {2014})}\BibitemShut {NoStop}%
\bibitem [{\citenamefont {Goold}\ \emph {et~al.}(2016)\citenamefont {Goold},
  \citenamefont {Huber}, \citenamefont {Riera}, \citenamefont {del Rio},\ and\
  \citenamefont {Skrzypczyk}}]{Huberreviewthermo2016}%
  \BibitemOpen
  \bibfield  {author} {\bibinfo {author} {\bibfnamefont {J.}~\bibnamefont
  {Goold}}, \bibinfo {author} {\bibfnamefont {M.}~\bibnamefont {Huber}},
  \bibinfo {author} {\bibfnamefont {A.}~\bibnamefont {Riera}}, \bibinfo
  {author} {\bibfnamefont {L.}~\bibnamefont {del Rio}}, \ and\ \bibinfo
  {author} {\bibfnamefont {P.}~\bibnamefont {Skrzypczyk}},\ }\href
  {http://stacks.iop.org/1751-8121/49/i=14/a=143001} {\bibfield  {journal}
  {\bibinfo  {journal} {Journal of Physics A: Mathematical and Theoretical}\
  }\textbf {\bibinfo {volume} {49}},\ \bibinfo {pages} {143001} (\bibinfo
  {year} {2016})}\BibitemShut {NoStop}%
\bibitem [{\citenamefont {Richens}\ and\ \citenamefont
  {Masanes}()}]{MasanesNatComm2016work}%
  \BibitemOpen
  \bibfield  {author} {\bibinfo {author} {\bibfnamefont {J.~G.}\ \bibnamefont
  {Richens}}\ and\ \bibinfo {author} {\bibfnamefont {L.}~\bibnamefont
  {Masanes}},\ }\href {\doibase 10.1038/ncomms13511} {\bibfield  {journal}
  {\bibinfo  {journal} {Nature communications}\ }\textbf {\bibinfo {volume}
  {7}},\ \bibinfo {pages} {13511}}\BibitemShut {NoStop}%
\bibitem [{\citenamefont {Masanes}\ and\ \citenamefont
  {Oppenheim}(2017)}]{masanes2017general}%
  \BibitemOpen
  \bibfield  {author} {\bibinfo {author} {\bibfnamefont {L.}~\bibnamefont
  {Masanes}}\ and\ \bibinfo {author} {\bibfnamefont {J.}~\bibnamefont
  {Oppenheim}},\ }\href {\doibase doi.org/10.1038/ncomms14538} {\bibfield
  {journal} {\bibinfo  {journal} {Nature communications}\ }\textbf {\bibinfo
  {volume} {8}},\ \bibinfo {pages} {14538} (\bibinfo {year}
  {2017})}\BibitemShut {NoStop}%
\bibitem [{\citenamefont {Alhambra}\ \emph {et~al.}(2019)\citenamefont
  {Alhambra}, \citenamefont {Styliaris}, \citenamefont
  {Rodr\'{\i}guez-Briones}, \citenamefont {Sikora},\ and\ \citenamefont
  {Mart\'{\i}n-Mart\'{\i}nez}}]{2019arXiv190202357A}%
  \BibitemOpen
  \bibfield  {author} {\bibinfo {author} {\bibfnamefont {A.~M.}\ \bibnamefont
  {Alhambra}}, \bibinfo {author} {\bibfnamefont {G.}~\bibnamefont {Styliaris}},
  \bibinfo {author} {\bibfnamefont {N.~A.}\ \bibnamefont
  {Rodr\'{\i}guez-Briones}}, \bibinfo {author} {\bibfnamefont {J.}~\bibnamefont
  {Sikora}}, \ and\ \bibinfo {author} {\bibfnamefont {E.}~\bibnamefont
  {Mart\'{\i}n-Mart\'{\i}nez}},\ }\href {\doibase
  10.1103/PhysRevLett.123.190601} {\bibfield  {journal} {\bibinfo  {journal}
  {Phys. Rev. Lett.}\ }\textbf {\bibinfo {volume} {123}},\ \bibinfo {pages}
  {190601} (\bibinfo {year} {2019})}\BibitemShut {NoStop}%
\bibitem [{\citenamefont {{\AA}berg}(2013)}]{aaberg2013truly}%
  \BibitemOpen
  \bibfield  {author} {\bibinfo {author} {\bibfnamefont {J.}~\bibnamefont
  {{\AA}berg}},\ }\href {\doibase doi.org/10.1038/ncomms2712} {\bibfield
  {journal} {\bibinfo  {journal} {Nature communications}\ }\textbf {\bibinfo
  {volume} {4}},\ \bibinfo {pages} {1925} (\bibinfo {year} {2013})}\BibitemShut
  {NoStop}%
\bibitem [{\citenamefont {{Frenzel}}\ \emph {et~al.}(2014)\citenamefont
  {{Frenzel}}, \citenamefont {{Jennings}},\ and\ \citenamefont
  {{Rudolph}}}]{FrenzelPRE2014}%
  \BibitemOpen
  \bibfield  {author} {\bibinfo {author} {\bibfnamefont {M.~F.}\ \bibnamefont
  {{Frenzel}}}, \bibinfo {author} {\bibfnamefont {D.}~\bibnamefont
  {{Jennings}}}, \ and\ \bibinfo {author} {\bibfnamefont {T.}~\bibnamefont
  {{Rudolph}}},\ }\href {\doibase 10.1103/PhysRevE.90.052136} {\bibfield
  {journal} {\bibinfo  {journal} {\pre}\ }\textbf {\bibinfo {volume} {90}},\
  \bibinfo {eid} {052136} (\bibinfo {year} {2014})}\BibitemShut {NoStop}%
\bibitem [{\citenamefont {Jarzynski}\ \emph {et~al.}(2015)\citenamefont
  {Jarzynski}, \citenamefont {Quan},\ and\ \citenamefont
  {Rahav}}]{jarzynski2015quantum}%
  \BibitemOpen
  \bibfield  {author} {\bibinfo {author} {\bibfnamefont {C.}~\bibnamefont
  {Jarzynski}}, \bibinfo {author} {\bibfnamefont {H.~T.}\ \bibnamefont {Quan}},
  \ and\ \bibinfo {author} {\bibfnamefont {S.}~\bibnamefont {Rahav}},\ }\href
  {\doibase 10.1103/PhysRevX.5.031038} {\bibfield  {journal} {\bibinfo
  {journal} {Phys. Rev. X}\ }\textbf {\bibinfo {volume} {5}},\ \bibinfo {pages}
  {031038} (\bibinfo {year} {2015})}\BibitemShut {NoStop}%
\bibitem [{\citenamefont {Alhambra}\ \emph {et~al.}(2016)\citenamefont
  {Alhambra}, \citenamefont {Masanes}, \citenamefont {Oppenheim},\ and\
  \citenamefont {Perry}}]{AlhambraPRX2016}%
  \BibitemOpen
  \bibfield  {author} {\bibinfo {author} {\bibfnamefont {A.~M.}\ \bibnamefont
  {Alhambra}}, \bibinfo {author} {\bibfnamefont {L.}~\bibnamefont {Masanes}},
  \bibinfo {author} {\bibfnamefont {J.}~\bibnamefont {Oppenheim}}, \ and\
  \bibinfo {author} {\bibfnamefont {C.}~\bibnamefont {Perry}},\ }\href
  {\doibase 10.1103/PhysRevX.6.041017} {\bibfield  {journal} {\bibinfo
  {journal} {Phys. Rev. X}\ }\textbf {\bibinfo {volume} {6}},\ \bibinfo {pages}
  {041017} (\bibinfo {year} {2016})}\BibitemShut {NoStop}%
\bibitem [{\citenamefont {Gallego}\ \emph {et~al.}(2016)\citenamefont
  {Gallego}, \citenamefont {Eisert},\ and\ \citenamefont
  {Wilming}}]{gallego2016thermodynamic}%
  \BibitemOpen
  \bibfield  {author} {\bibinfo {author} {\bibfnamefont {R.}~\bibnamefont
  {Gallego}}, \bibinfo {author} {\bibfnamefont {J.}~\bibnamefont {Eisert}}, \
  and\ \bibinfo {author} {\bibfnamefont {H.}~\bibnamefont {Wilming}},\ }\href
  {\doibase 10.1088/1367-2630/18/10/103017} {\bibfield  {journal} {\bibinfo
  {journal} {New Journal of Physics}\ }\textbf {\bibinfo {volume} {18}},\
  \bibinfo {pages} {103017} (\bibinfo {year} {2016})}\BibitemShut {NoStop}%
\bibitem [{\citenamefont {Brandner}\ \emph {et~al.}(2015)\citenamefont
  {Brandner}, \citenamefont {Bauer}, \citenamefont {Schmid},\ and\
  \citenamefont {Seifert}}]{brandner2015coherence}%
  \BibitemOpen
  \bibfield  {author} {\bibinfo {author} {\bibfnamefont {K.}~\bibnamefont
  {Brandner}}, \bibinfo {author} {\bibfnamefont {M.}~\bibnamefont {Bauer}},
  \bibinfo {author} {\bibfnamefont {M.~T.}\ \bibnamefont {Schmid}}, \ and\
  \bibinfo {author} {\bibfnamefont {U.}~\bibnamefont {Seifert}},\ }\href
  {\doibase 10.1088/1367-2630/17/6/065006} {\bibfield  {journal} {\bibinfo
  {journal} {New Journal of Physics}\ }\textbf {\bibinfo {volume} {17}},\
  \bibinfo {pages} {065006} (\bibinfo {year} {2015})}\BibitemShut {NoStop}%
\bibitem [{\citenamefont {Lostaglio}\ \emph {et~al.}(2015)\citenamefont
  {Lostaglio}, \citenamefont {Korzekwa}, \citenamefont {Jennings},\ and\
  \citenamefont {Rudolph}}]{lostaglio2015quantum}%
  \BibitemOpen
  \bibfield  {author} {\bibinfo {author} {\bibfnamefont {M.}~\bibnamefont
  {Lostaglio}}, \bibinfo {author} {\bibfnamefont {K.}~\bibnamefont {Korzekwa}},
  \bibinfo {author} {\bibfnamefont {D.}~\bibnamefont {Jennings}}, \ and\
  \bibinfo {author} {\bibfnamefont {T.}~\bibnamefont {Rudolph}},\ }\href
  {\doibase 10.1103/PhysRevX.5.021001} {\bibfield  {journal} {\bibinfo
  {journal} {Phys. Rev. X}\ }\textbf {\bibinfo {volume} {5}},\ \bibinfo {pages}
  {021001} (\bibinfo {year} {2015})}\BibitemShut {NoStop}%
\bibitem [{\citenamefont {Uzdin}\ \emph {et~al.}(2015)\citenamefont {Uzdin},
  \citenamefont {Levy},\ and\ \citenamefont {Kosloff}}]{UzdinPRX2015}%
  \BibitemOpen
  \bibfield  {author} {\bibinfo {author} {\bibfnamefont {R.}~\bibnamefont
  {Uzdin}}, \bibinfo {author} {\bibfnamefont {A.}~\bibnamefont {Levy}}, \ and\
  \bibinfo {author} {\bibfnamefont {R.}~\bibnamefont {Kosloff}},\ }\href
  {\doibase 10.1103/PhysRevX.5.031044} {\bibfield  {journal} {\bibinfo
  {journal} {Phys. Rev. X}\ }\textbf {\bibinfo {volume} {5}},\ \bibinfo {pages}
  {031044} (\bibinfo {year} {2015})}\BibitemShut {NoStop}%
\bibitem [{\citenamefont {Korzekwa}\ \emph {et~al.}(2016)\citenamefont
  {Korzekwa}, \citenamefont {Lostaglio}, \citenamefont {Oppenheim},\ and\
  \citenamefont {Jennings}}]{korzekwa2016extraction}%
  \BibitemOpen
  \bibfield  {author} {\bibinfo {author} {\bibfnamefont {K.}~\bibnamefont
  {Korzekwa}}, \bibinfo {author} {\bibfnamefont {M.}~\bibnamefont {Lostaglio}},
  \bibinfo {author} {\bibfnamefont {J.}~\bibnamefont {Oppenheim}}, \ and\
  \bibinfo {author} {\bibfnamefont {D.}~\bibnamefont {Jennings}},\ }\href
  {\doibase 10.1088/1367-2630/18/2/023045} {\bibfield  {journal} {\bibinfo
  {journal} {New Journal of Physics}\ }\textbf {\bibinfo {volume} {18}},\
  \bibinfo {pages} {023045} (\bibinfo {year} {2016})}\BibitemShut {NoStop}%
\bibitem [{\citenamefont {{Streltsov}}\ \emph {et~al.}(2017)\citenamefont
  {{Streltsov}}, \citenamefont {{Adesso}},\ and\ \citenamefont
  {{Plenio}}}]{CoherenceRevMod2017}%
  \BibitemOpen
  \bibfield  {author} {\bibinfo {author} {\bibfnamefont {A.}~\bibnamefont
  {{Streltsov}}}, \bibinfo {author} {\bibfnamefont {G.}~\bibnamefont
  {{Adesso}}}, \ and\ \bibinfo {author} {\bibfnamefont {M.~B.}\ \bibnamefont
  {{Plenio}}},\ }\href {\doibase 10.1103/RevModPhys.89.041003} {\bibfield
  {journal} {\bibinfo  {journal} {Reviews of Modern Physics}\ }\textbf
  {\bibinfo {volume} {89}},\ \bibinfo {eid} {041003} (\bibinfo {year}
  {2017})}\BibitemShut {NoStop}%
\bibitem [{\citenamefont {Francica}\ \emph {et~al.}(2019)\citenamefont
  {Francica}, \citenamefont {Goold},\ and\ \citenamefont
  {Plastina}}]{FrancicaPRE2019}%
  \BibitemOpen
  \bibfield  {author} {\bibinfo {author} {\bibfnamefont {G.}~\bibnamefont
  {Francica}}, \bibinfo {author} {\bibfnamefont {J.}~\bibnamefont {Goold}}, \
  and\ \bibinfo {author} {\bibfnamefont {F.}~\bibnamefont {Plastina}},\ }\href
  {\doibase 10.1103/PhysRevE.99.042105} {\bibfield  {journal} {\bibinfo
  {journal} {Phys. Rev. E}\ }\textbf {\bibinfo {volume} {99}},\ \bibinfo
  {pages} {042105} (\bibinfo {year} {2019})}\BibitemShut {NoStop}%
\bibitem [{\citenamefont {{Latune}}\ \emph {et~al.}(2019)\citenamefont
  {{Latune}}, \citenamefont {{Sinayskiy}},\ and\ \citenamefont
  {{Petruccione}}}]{PetruccioneSciRep2019}%
  \BibitemOpen
  \bibfield  {author} {\bibinfo {author} {\bibfnamefont {C.~L.}\ \bibnamefont
  {{Latune}}}, \bibinfo {author} {\bibfnamefont {I.}~\bibnamefont
  {{Sinayskiy}}}, \ and\ \bibinfo {author} {\bibfnamefont {F.}~\bibnamefont
  {{Petruccione}}},\ }\href {\doibase 10.1038/s41598-019-39300-4} {\bibfield
  {journal} {\bibinfo  {journal} {Scientific Reports}\ }\textbf {\bibinfo
  {volume} {9}},\ \bibinfo {eid} {3191} (\bibinfo {year} {2019})}\BibitemShut
  {NoStop}%
\bibitem [{\citenamefont {Andolina}\ \emph
  {et~al.}(2019{\natexlab{a}})\citenamefont {Andolina}, \citenamefont {Keck},
  \citenamefont {Mari}, \citenamefont {Campisi}, \citenamefont {Giovannetti},\
  and\ \citenamefont {Polini}}]{PoliniPRL2019}%
  \BibitemOpen
  \bibfield  {author} {\bibinfo {author} {\bibfnamefont {G.~M.}\ \bibnamefont
  {Andolina}}, \bibinfo {author} {\bibfnamefont {M.}~\bibnamefont {Keck}},
  \bibinfo {author} {\bibfnamefont {A.}~\bibnamefont {Mari}}, \bibinfo {author}
  {\bibfnamefont {M.}~\bibnamefont {Campisi}}, \bibinfo {author} {\bibfnamefont
  {V.}~\bibnamefont {Giovannetti}}, \ and\ \bibinfo {author} {\bibfnamefont
  {M.}~\bibnamefont {Polini}},\ }\href {\doibase
  10.1103/PhysRevLett.122.047702} {\bibfield  {journal} {\bibinfo  {journal}
  {Phys. Rev. Lett.}\ }\textbf {\bibinfo {volume} {122}},\ \bibinfo {pages}
  {047702} (\bibinfo {year} {2019}{\natexlab{a}})}\BibitemShut {NoStop}%
\bibitem [{\citenamefont {{Brunner}}\ \emph {et~al.}(2012)\citenamefont
  {{Brunner}}, \citenamefont {{Linden}}, \citenamefont {{Popescu}},\ and\
  \citenamefont {{Skrzypczyk}}}]{BrunnerPRE2012}%
  \BibitemOpen
  \bibfield  {author} {\bibinfo {author} {\bibfnamefont {N.}~\bibnamefont
  {{Brunner}}}, \bibinfo {author} {\bibfnamefont {N.}~\bibnamefont {{Linden}}},
  \bibinfo {author} {\bibfnamefont {S.}~\bibnamefont {{Popescu}}}, \ and\
  \bibinfo {author} {\bibfnamefont {P.}~\bibnamefont {{Skrzypczyk}}},\ }\href
  {\doibase 10.1103/PhysRevE.85.051117} {\bibfield  {journal} {\bibinfo
  {journal} {\pre}\ }\textbf {\bibinfo {volume} {85}},\ \bibinfo {eid} {051117}
  (\bibinfo {year} {2012})}\BibitemShut {NoStop}%
\bibitem [{\citenamefont {Linden}\ \emph {et~al.}(2010)\citenamefont {Linden},
  \citenamefont {Popescu},\ and\ \citenamefont {Skrzypczyk}}]{LindenPRL2010}%
  \BibitemOpen
  \bibfield  {author} {\bibinfo {author} {\bibfnamefont {N.}~\bibnamefont
  {Linden}}, \bibinfo {author} {\bibfnamefont {S.}~\bibnamefont {Popescu}}, \
  and\ \bibinfo {author} {\bibfnamefont {P.}~\bibnamefont {Skrzypczyk}},\
  }\href {\doibase 10.1103/PhysRevLett.105.130401} {\bibfield  {journal}
  {\bibinfo  {journal} {Phys. Rev. Lett.}\ }\textbf {\bibinfo {volume} {105}},\
  \bibinfo {pages} {130401} (\bibinfo {year} {2010})}\BibitemShut {NoStop}%
\bibitem [{\citenamefont {Correa}\ \emph {et~al.}(2013)\citenamefont {Correa},
  \citenamefont {Palao}, \citenamefont {Adesso},\ and\ \citenamefont
  {Alonso}}]{CorreaPRE2013}%
  \BibitemOpen
  \bibfield  {author} {\bibinfo {author} {\bibfnamefont {L.~A.}\ \bibnamefont
  {Correa}}, \bibinfo {author} {\bibfnamefont {J.~P.}\ \bibnamefont {Palao}},
  \bibinfo {author} {\bibfnamefont {G.}~\bibnamefont {Adesso}}, \ and\ \bibinfo
  {author} {\bibfnamefont {D.}~\bibnamefont {Alonso}},\ }\href {\doibase
  10.1103/PhysRevE.87.042131} {\bibfield  {journal} {\bibinfo  {journal} {Phys.
  Rev. E}\ }\textbf {\bibinfo {volume} {87}},\ \bibinfo {pages} {042131}
  (\bibinfo {year} {2013})}\BibitemShut {NoStop}%
\bibitem [{\citenamefont {{Skrzypczyk}}\ \emph
  {et~al.}(2014{\natexlab{a}})\citenamefont {{Skrzypczyk}}, \citenamefont
  {{Short}},\ and\ \citenamefont {{Popescu}}}]{PopescuNatComm2014}%
  \BibitemOpen
  \bibfield  {author} {\bibinfo {author} {\bibfnamefont {P.}~\bibnamefont
  {{Skrzypczyk}}}, \bibinfo {author} {\bibfnamefont {A.~J.}\ \bibnamefont
  {{Short}}}, \ and\ \bibinfo {author} {\bibfnamefont {S.}~\bibnamefont
  {{Popescu}}},\ }\href {\doibase 10.1038/ncomms5185} {\bibfield  {journal}
  {\bibinfo  {journal} {Nature Communications}\ }\textbf {\bibinfo {volume}
  {5}},\ \bibinfo {eid} {4185} (\bibinfo {year}
  {2014}{\natexlab{a}})}\BibitemShut {NoStop}%
\bibitem [{\citenamefont {{Kosloff}}\ and\ \citenamefont
  {{Levy}}(2014)}]{Kosloff2014}%
  \BibitemOpen
  \bibfield  {author} {\bibinfo {author} {\bibfnamefont {R.}~\bibnamefont
  {{Kosloff}}}\ and\ \bibinfo {author} {\bibfnamefont {A.}~\bibnamefont
  {{Levy}}},\ }\href {\doibase 10.1146/annurev-physchem-040513-103724}
  {\bibfield  {journal} {\bibinfo  {journal} {Annual Review of Physical
  Chemistry}\ }\textbf {\bibinfo {volume} {65}},\ \bibinfo {pages} {365}
  (\bibinfo {year} {2014})}\BibitemShut {NoStop}%
\bibitem [{\citenamefont {del Campo}\ \emph {et~al.}(2014)\citenamefont {del
  Campo}, \citenamefont {Goold},\ and\ \citenamefont {Paternostro}}]{AdC2014}%
  \BibitemOpen
  \bibfield  {author} {\bibinfo {author} {\bibfnamefont {A.}~\bibnamefont {del
  Campo}}, \bibinfo {author} {\bibfnamefont {J.}~\bibnamefont {Goold}}, \ and\
  \bibinfo {author} {\bibfnamefont {M.}~\bibnamefont {Paternostro}},\ }\href
  {\doibase 10.1038/srep06208} {\bibfield  {journal} {\bibinfo  {journal}
  {Scientific reports}\ }\textbf {\bibinfo {volume} {4}} (\bibinfo {year}
  {2014}),\ 10.1038/srep06208}\BibitemShut {NoStop}%
\bibitem [{\citenamefont {Alicki}\ and\ \citenamefont
  {Gelbwaser-Klimovsky}(2015)}]{GelbwaserNJP2015}%
  \BibitemOpen
  \bibfield  {author} {\bibinfo {author} {\bibfnamefont {R.}~\bibnamefont
  {Alicki}}\ and\ \bibinfo {author} {\bibfnamefont {D.}~\bibnamefont
  {Gelbwaser-Klimovsky}},\ }\href
  {http://stacks.iop.org/1367-2630/17/i=11/a=115012} {\bibfield  {journal}
  {\bibinfo  {journal} {New Journal of Physics}\ }\textbf {\bibinfo {volume}
  {17}},\ \bibinfo {pages} {115012} (\bibinfo {year} {2015})}\BibitemShut
  {NoStop}%
\bibitem [{\citenamefont {Hofer}\ \emph {et~al.}(2016)\citenamefont {Hofer},
  \citenamefont {Perarnau-Llobet}, \citenamefont {Brask}, \citenamefont
  {Silva}, \citenamefont {Huber},\ and\ \citenamefont
  {Brunner}}]{BrunnerPRB2016}%
  \BibitemOpen
  \bibfield  {author} {\bibinfo {author} {\bibfnamefont {P.~P.}\ \bibnamefont
  {Hofer}}, \bibinfo {author} {\bibfnamefont {M.}~\bibnamefont
  {Perarnau-Llobet}}, \bibinfo {author} {\bibfnamefont {J.~B.}\ \bibnamefont
  {Brask}}, \bibinfo {author} {\bibfnamefont {R.}~\bibnamefont {Silva}},
  \bibinfo {author} {\bibfnamefont {M.}~\bibnamefont {Huber}}, \ and\ \bibinfo
  {author} {\bibfnamefont {N.}~\bibnamefont {Brunner}},\ }\href {\doibase
  10.1103/PhysRevB.94.235420} {\bibfield  {journal} {\bibinfo  {journal} {Phys.
  Rev. B}\ }\textbf {\bibinfo {volume} {94}},\ \bibinfo {pages} {235420}
  (\bibinfo {year} {2016})}\BibitemShut {NoStop}%
\bibitem [{\citenamefont {Niedenzu}\ \emph {et~al.}(2016)\citenamefont
  {Niedenzu}, \citenamefont {Gelbwaser-Klimovsky}, \citenamefont {Kofman},\
  and\ \citenamefont {Kurizki}}]{GelbwaserNJP2016}%
  \BibitemOpen
  \bibfield  {author} {\bibinfo {author} {\bibfnamefont {W.}~\bibnamefont
  {Niedenzu}}, \bibinfo {author} {\bibfnamefont {D.}~\bibnamefont
  {Gelbwaser-Klimovsky}}, \bibinfo {author} {\bibfnamefont {A.~G.}\
  \bibnamefont {Kofman}}, \ and\ \bibinfo {author} {\bibfnamefont
  {G.}~\bibnamefont {Kurizki}},\ }\href
  {http://stacks.iop.org/1367-2630/18/i=8/a=083012} {\bibfield  {journal}
  {\bibinfo  {journal} {New Journal of Physics}\ }\textbf {\bibinfo {volume}
  {18}},\ \bibinfo {pages} {083012} (\bibinfo {year} {2016})}\BibitemShut
  {NoStop}%
\bibitem [{\citenamefont {Caravelli}\ \emph {et~al.}(2020)\citenamefont
  {Caravelli}, \citenamefont {Coulter-De~Wit}, \citenamefont
  {Garc\'{\i}a-Pintos},\ and\ \citenamefont {Hamma}}]{caravelli2019random}%
  \BibitemOpen
  \bibfield  {author} {\bibinfo {author} {\bibfnamefont {F.}~\bibnamefont
  {Caravelli}}, \bibinfo {author} {\bibfnamefont {G.}~\bibnamefont
  {Coulter-De~Wit}}, \bibinfo {author} {\bibfnamefont {L.~P.}\ \bibnamefont
  {Garc\'{\i}a-Pintos}}, \ and\ \bibinfo {author} {\bibfnamefont
  {A.}~\bibnamefont {Hamma}},\ }\href {\doibase
  10.1103/PhysRevResearch.2.023095} {\bibfield  {journal} {\bibinfo  {journal}
  {Phys. Rev. Research}\ }\textbf {\bibinfo {volume} {2}},\ \bibinfo {pages}
  {023095} (\bibinfo {year} {2020})}\BibitemShut {NoStop}%
\bibitem [{\citenamefont {Funo}\ and\ \citenamefont
  {Ueda}(2015)}]{Masahito2015}%
  \BibitemOpen
  \bibfield  {author} {\bibinfo {author} {\bibfnamefont {K.}~\bibnamefont
  {Funo}}\ and\ \bibinfo {author} {\bibfnamefont {M.}~\bibnamefont {Ueda}},\
  }\href {\doibase 10.1103/PhysRevLett.115.260601} {\bibfield  {journal}
  {\bibinfo  {journal} {Phys. Rev. Lett.}\ }\textbf {\bibinfo {volume} {115}},\
  \bibinfo {pages} {260601} (\bibinfo {year} {2015})}\BibitemShut {NoStop}%
\bibitem [{\citenamefont {Funo}\ \emph {et~al.}(2017)\citenamefont {Funo},
  \citenamefont {Zhang}, \citenamefont {Chatou}, \citenamefont {Kim},
  \citenamefont {Ueda},\ and\ \citenamefont {del Campo}}]{Funo17}%
  \BibitemOpen
  \bibfield  {author} {\bibinfo {author} {\bibfnamefont {K.}~\bibnamefont
  {Funo}}, \bibinfo {author} {\bibfnamefont {J.-N.}\ \bibnamefont {Zhang}},
  \bibinfo {author} {\bibfnamefont {C.}~\bibnamefont {Chatou}}, \bibinfo
  {author} {\bibfnamefont {K.}~\bibnamefont {Kim}}, \bibinfo {author}
  {\bibfnamefont {M.}~\bibnamefont {Ueda}}, \ and\ \bibinfo {author}
  {\bibfnamefont {A.}~\bibnamefont {del Campo}},\ }\href {\doibase
  10.1103/PhysRevLett.118.100602} {\bibfield  {journal} {\bibinfo  {journal}
  {Phys. Rev. Lett.}\ }\textbf {\bibinfo {volume} {118}},\ \bibinfo {pages}
  {100602} (\bibinfo {year} {2017})}\BibitemShut {NoStop}%
\bibitem [{\citenamefont {Binder}\ \emph {et~al.}(2015)\citenamefont {Binder},
  \citenamefont {Vinjanampathy}, \citenamefont {Modi},\ and\ \citenamefont
  {Goold}}]{Binder15}%
  \BibitemOpen
  \bibfield  {author} {\bibinfo {author} {\bibfnamefont {F.~C.}\ \bibnamefont
  {Binder}}, \bibinfo {author} {\bibfnamefont {S.}~\bibnamefont
  {Vinjanampathy}}, \bibinfo {author} {\bibfnamefont {K.}~\bibnamefont {Modi}},
  \ and\ \bibinfo {author} {\bibfnamefont {J.}~\bibnamefont {Goold}},\ }\href
  {\doibase 10.1088/1367-2630/17/7/075015} {\bibfield  {journal} {\bibinfo
  {journal} {New Journal of Physics}\ }\textbf {\bibinfo {volume} {17}},\
  \bibinfo {pages} {075015} (\bibinfo {year} {2015})}\BibitemShut {NoStop}%
\bibitem [{\citenamefont {Jaramillo}\ \emph {et~al.}(2016)\citenamefont
  {Jaramillo}, \citenamefont {Beau},\ and\ \citenamefont {del
  Campo}}]{AdCNJP2016}%
  \BibitemOpen
  \bibfield  {author} {\bibinfo {author} {\bibfnamefont {J.}~\bibnamefont
  {Jaramillo}}, \bibinfo {author} {\bibfnamefont {M.}~\bibnamefont {Beau}}, \
  and\ \bibinfo {author} {\bibfnamefont {A.}~\bibnamefont {del Campo}},\ }\href
  {http://stacks.iop.org/1367-2630/18/i=7/a=075019} {\bibfield  {journal}
  {\bibinfo  {journal} {New Journal of Physics}\ }\textbf {\bibinfo {volume}
  {18}},\ \bibinfo {pages} {075019} (\bibinfo {year} {2016})}\BibitemShut
  {NoStop}%
\bibitem [{\citenamefont {Campaioli}\ \emph {et~al.}(2017)\citenamefont
  {Campaioli}, \citenamefont {Pollock}, \citenamefont {Binder}, \citenamefont
  {C\'eleri}, \citenamefont {Goold}, \citenamefont {Vinjanampathy},\ and\
  \citenamefont {Modi}}]{ModiPRL2017}%
  \BibitemOpen
  \bibfield  {author} {\bibinfo {author} {\bibfnamefont {F.}~\bibnamefont
  {Campaioli}}, \bibinfo {author} {\bibfnamefont {F.~A.}\ \bibnamefont
  {Pollock}}, \bibinfo {author} {\bibfnamefont {F.~C.}\ \bibnamefont {Binder}},
  \bibinfo {author} {\bibfnamefont {L.}~\bibnamefont {C\'eleri}}, \bibinfo
  {author} {\bibfnamefont {J.}~\bibnamefont {Goold}}, \bibinfo {author}
  {\bibfnamefont {S.}~\bibnamefont {Vinjanampathy}}, \ and\ \bibinfo {author}
  {\bibfnamefont {K.}~\bibnamefont {Modi}},\ }\href {\doibase
  10.1103/PhysRevLett.118.150601} {\bibfield  {journal} {\bibinfo  {journal}
  {Phys. Rev. Lett.}\ }\textbf {\bibinfo {volume} {118}},\ \bibinfo {pages}
  {150601} (\bibinfo {year} {2017})}\BibitemShut {NoStop}%
\bibitem [{\citenamefont {Ferraro}\ \emph {et~al.}(2018)\citenamefont
  {Ferraro}, \citenamefont {Campisi}, \citenamefont {Andolina}, \citenamefont
  {Pellegrini},\ and\ \citenamefont {Polini}}]{FerraroPRL2018}%
  \BibitemOpen
  \bibfield  {author} {\bibinfo {author} {\bibfnamefont {D.}~\bibnamefont
  {Ferraro}}, \bibinfo {author} {\bibfnamefont {M.}~\bibnamefont {Campisi}},
  \bibinfo {author} {\bibfnamefont {G.~M.}\ \bibnamefont {Andolina}}, \bibinfo
  {author} {\bibfnamefont {V.}~\bibnamefont {Pellegrini}}, \ and\ \bibinfo
  {author} {\bibfnamefont {M.}~\bibnamefont {Polini}},\ }\href {\doibase
  10.1103/PhysRevLett.120.117702} {\bibfield  {journal} {\bibinfo  {journal}
  {Phys. Rev. Lett.}\ }\textbf {\bibinfo {volume} {120}},\ \bibinfo {pages}
  {117702} (\bibinfo {year} {2018})}\BibitemShut {NoStop}%
\bibitem [{\citenamefont {Perarnau-Llobet}\ and\ \citenamefont
  {Uzdin}(2019)}]{Perarnau_Llobet_2019}%
  \BibitemOpen
  \bibfield  {author} {\bibinfo {author} {\bibfnamefont {M.}~\bibnamefont
  {Perarnau-Llobet}}\ and\ \bibinfo {author} {\bibfnamefont {R.}~\bibnamefont
  {Uzdin}},\ }\href {\doibase 10.1088/1367-2630/ab36a9} {\bibfield  {journal}
  {\bibinfo  {journal} {New Journal of Physics}\ }\textbf {\bibinfo {volume}
  {21}},\ \bibinfo {pages} {083023} (\bibinfo {year} {2019})}\BibitemShut
  {NoStop}%
\bibitem [{\citenamefont {Watanabe}\ \emph {et~al.}(2017)\citenamefont
  {Watanabe}, \citenamefont {Venkatesh}, \citenamefont {Talkner},\ and\
  \citenamefont {del Campo}}]{AdCPRL2017cycleengine}%
  \BibitemOpen
  \bibfield  {author} {\bibinfo {author} {\bibfnamefont {G.}~\bibnamefont
  {Watanabe}}, \bibinfo {author} {\bibfnamefont {B.~P.}\ \bibnamefont
  {Venkatesh}}, \bibinfo {author} {\bibfnamefont {P.}~\bibnamefont {Talkner}},
  \ and\ \bibinfo {author} {\bibfnamefont {A.}~\bibnamefont {del Campo}},\
  }\href {\doibase 10.1103/PhysRevLett.118.050601} {\bibfield  {journal}
  {\bibinfo  {journal} {Phys. Rev. Lett.}\ }\textbf {\bibinfo {volume} {118}},\
  \bibinfo {pages} {050601} (\bibinfo {year} {2017})}\BibitemShut {NoStop}%
\bibitem [{\citenamefont {{Watanabe}}\ \emph {et~al.}(2019)\citenamefont
  {{Watanabe}}, \citenamefont {{Venkatesh}}, \citenamefont {{Hwang}},\ and\
  \citenamefont {{del Campo}}}]{Watanabe19}%
  \BibitemOpen
  \bibfield  {author} {\bibinfo {author} {\bibfnamefont {G.}~\bibnamefont
  {{Watanabe}}}, \bibinfo {author} {\bibfnamefont {P.}~\bibnamefont
  {{Venkatesh}}, \bibfnamefont {B.~P.and~{Talkner}}}, \bibinfo {author}
  {\bibfnamefont {M.-J.}\ \bibnamefont {{Hwang}}}, \ and\ \bibinfo {author}
  {\bibfnamefont {A.}~\bibnamefont {{del Campo}}},\ }\href@noop {} {\bibfield
  {journal} {\bibinfo  {journal} {arXiv e-prints}\ ,\ \bibinfo {eid}
  {arXiv:1904.07811}} (\bibinfo {year} {2019})},\ \Eprint
  {http://arxiv.org/abs/1904.07811} {arXiv:1904.07811 [quant-ph]} \BibitemShut
  {NoStop}%
\bibitem [{\citenamefont
  {M\"uller}(2018{\natexlab{a}})}]{MuellerCorrelatedMachines2017}%
  \BibitemOpen
  \bibfield  {author} {\bibinfo {author} {\bibfnamefont {M.~P.}\ \bibnamefont
  {M\"uller}},\ }\href {\doibase 10.1103/PhysRevX.8.041051} {\bibfield
  {journal} {\bibinfo  {journal} {Phys. Rev. X}\ }\textbf {\bibinfo {volume}
  {8}},\ \bibinfo {pages} {041051} (\bibinfo {year}
  {2018}{\natexlab{a}})}\BibitemShut {NoStop}%
\bibitem [{\citenamefont {Hovhannisyan}\ \emph {et~al.}(2013)\citenamefont
  {Hovhannisyan}, \citenamefont {Perarnau-Llobet}, \citenamefont {Huber},\ and\
  \citenamefont {Ac\'{\i}n}}]{KarenPRL2013}%
  \BibitemOpen
  \bibfield  {author} {\bibinfo {author} {\bibfnamefont {K.~V.}\ \bibnamefont
  {Hovhannisyan}}, \bibinfo {author} {\bibfnamefont {M.}~\bibnamefont
  {Perarnau-Llobet}}, \bibinfo {author} {\bibfnamefont {M.}~\bibnamefont
  {Huber}}, \ and\ \bibinfo {author} {\bibfnamefont {A.}~\bibnamefont
  {Ac\'{\i}n}},\ }\href {\doibase 10.1103/PhysRevLett.111.240401} {\bibfield
  {journal} {\bibinfo  {journal} {Phys. Rev. Lett.}\ }\textbf {\bibinfo
  {volume} {111}},\ \bibinfo {pages} {240401} (\bibinfo {year}
  {2013})}\BibitemShut {NoStop}%
\bibitem [{\citenamefont {Friis}\ and\ \citenamefont
  {Huber}(2018)}]{HuberGaussianBatteries2017}%
  \BibitemOpen
  \bibfield  {author} {\bibinfo {author} {\bibfnamefont {N.}~\bibnamefont
  {Friis}}\ and\ \bibinfo {author} {\bibfnamefont {M.}~\bibnamefont {Huber}},\
  }\href {\doibase 10.22331/q-2018-04-23-61} {\bibfield  {journal} {\bibinfo
  {journal} {{Quantum}}\ }\textbf {\bibinfo {volume} {2}},\ \bibinfo {pages}
  {61} (\bibinfo {year} {2018})}\BibitemShut {NoStop}%
\bibitem [{\citenamefont {Mandelstam}\ and\ \citenamefont
  {Tamm}(1945)}]{mandelstamtamm1945}%
  \BibitemOpen
  \bibfield  {author} {\bibinfo {author} {\bibfnamefont {L.}~\bibnamefont
  {Mandelstam}}\ and\ \bibinfo {author} {\bibfnamefont {I.}~\bibnamefont
  {Tamm}},\ }\href@noop {} {\bibfield  {journal} {\bibinfo  {journal} {J.
  Phys.(USSR)}\ }\textbf {\bibinfo {volume} {9}},\ \bibinfo {pages} {1}
  (\bibinfo {year} {1945})}\BibitemShut {NoStop}%
\bibitem [{\citenamefont {Anandan}\ and\ \citenamefont
  {Aharonov}(1990)}]{YakirPRL1990}%
  \BibitemOpen
  \bibfield  {author} {\bibinfo {author} {\bibfnamefont {J.}~\bibnamefont
  {Anandan}}\ and\ \bibinfo {author} {\bibfnamefont {Y.}~\bibnamefont
  {Aharonov}},\ }\href {\doibase 10.1103/PhysRevLett.65.1697} {\bibfield
  {journal} {\bibinfo  {journal} {Phys. Rev. Lett.}\ }\textbf {\bibinfo
  {volume} {65}},\ \bibinfo {pages} {1697} (\bibinfo {year}
  {1990})}\BibitemShut {NoStop}%
\bibitem [{\citenamefont {Margolus}\ and\ \citenamefont
  {Levitin}(1998)}]{margoluslevitin1998}%
  \BibitemOpen
  \bibfield  {author} {\bibinfo {author} {\bibfnamefont {N.}~\bibnamefont
  {Margolus}}\ and\ \bibinfo {author} {\bibfnamefont {L.~B.}\ \bibnamefont
  {Levitin}},\ }\href {\doibase https://doi.org/10.1016/S0167-2789(98)00054-2}
  {\bibfield  {journal} {\bibinfo  {journal} {Physica D: Nonlinear Phenomena}\
  }\textbf {\bibinfo {volume} {120}},\ \bibinfo {pages} {188 } (\bibinfo {year}
  {1998})},\ \bibinfo {note} {proceedings of the Fourth Workshop on Physics and
  Consumption}\BibitemShut {NoStop}%
\bibitem [{\citenamefont {Lloyd}(2000)}]{LloydNature2000}%
  \BibitemOpen
  \bibfield  {author} {\bibinfo {author} {\bibfnamefont {S.}~\bibnamefont
  {Lloyd}},\ }\href {\doibase 10.1038/35023282} {\bibfield  {journal} {\bibinfo
   {journal} {Nature}\ }\textbf {\bibinfo {volume} {406}},\ \bibinfo {pages}
  {1047} (\bibinfo {year} {2000})}\BibitemShut {NoStop}%
\bibitem [{\citenamefont {Lloyd}(2002)}]{LloydPRL2002}%
  \BibitemOpen
  \bibfield  {author} {\bibinfo {author} {\bibfnamefont {S.}~\bibnamefont
  {Lloyd}},\ }\href {\doibase 10.1103/PhysRevLett.88.237901} {\bibfield
  {journal} {\bibinfo  {journal} {Phys. Rev. Lett.}\ }\textbf {\bibinfo
  {volume} {88}},\ \bibinfo {pages} {237901} (\bibinfo {year}
  {2002})}\BibitemShut {NoStop}%
\bibitem [{\citenamefont {Giovannetti}\ \emph {et~al.}(2003)\citenamefont
  {Giovannetti}, \citenamefont {Lloyd},\ and\ \citenamefont
  {Maccone}}]{MacconePRA2003}%
  \BibitemOpen
  \bibfield  {author} {\bibinfo {author} {\bibfnamefont {V.}~\bibnamefont
  {Giovannetti}}, \bibinfo {author} {\bibfnamefont {S.}~\bibnamefont {Lloyd}},
  \ and\ \bibinfo {author} {\bibfnamefont {L.}~\bibnamefont {Maccone}},\ }\href
  {\doibase 10.1103/PhysRevA.67.052109} {\bibfield  {journal} {\bibinfo
  {journal} {Phys. Rev. A}\ }\textbf {\bibinfo {volume} {67}},\ \bibinfo
  {pages} {052109} (\bibinfo {year} {2003})}\BibitemShut {NoStop}%
\bibitem [{\citenamefont {{Taddei}}\ \emph {et~al.}(2013)\citenamefont
  {{Taddei}}, \citenamefont {{Escher}}, \citenamefont {{Davidovich}},\ and\
  \citenamefont {{de Matos Filho}}}]{DavidovichPRL2013}%
  \BibitemOpen
  \bibfield  {author} {\bibinfo {author} {\bibfnamefont {M.~M.}\ \bibnamefont
  {{Taddei}}}, \bibinfo {author} {\bibfnamefont {B.~M.}\ \bibnamefont
  {{Escher}}}, \bibinfo {author} {\bibfnamefont {L.}~\bibnamefont
  {{Davidovich}}}, \ and\ \bibinfo {author} {\bibfnamefont {R.~L.}\
  \bibnamefont {{de Matos Filho}}},\ }\href {\doibase
  10.1103/PhysRevLett.110.050402} {\bibfield  {journal} {\bibinfo  {journal}
  {Physical Review Letters}\ }\textbf {\bibinfo {volume} {110}},\ \bibinfo
  {eid} {050402} (\bibinfo {year} {2013})}\BibitemShut {NoStop}%
\bibitem [{\citenamefont {{del Campo}}\ \emph {et~al.}(2013)\citenamefont {{del
  Campo}}, \citenamefont {{Egusquiza}}, \citenamefont {{Plenio}},\ and\
  \citenamefont {{Huelga}}}]{delCampoPRL2013}%
  \BibitemOpen
  \bibfield  {author} {\bibinfo {author} {\bibfnamefont {A.}~\bibnamefont {{del
  Campo}}}, \bibinfo {author} {\bibfnamefont {I.~L.}\ \bibnamefont
  {{Egusquiza}}}, \bibinfo {author} {\bibfnamefont {M.~B.}\ \bibnamefont
  {{Plenio}}}, \ and\ \bibinfo {author} {\bibfnamefont {S.~F.}\ \bibnamefont
  {{Huelga}}},\ }\href {\doibase 10.1103/PhysRevLett.110.050403} {\bibfield
  {journal} {\bibinfo  {journal} {Physical Review Letters}\ }\textbf {\bibinfo
  {volume} {110}},\ \bibinfo {eid} {050403} (\bibinfo {year}
  {2013})}\BibitemShut {NoStop}%
\bibitem [{\citenamefont {{Deffner}}\ and\ \citenamefont
  {{Lutz}}(2013)}]{DeffnerLutzPRL2013}%
  \BibitemOpen
  \bibfield  {author} {\bibinfo {author} {\bibfnamefont {S.}~\bibnamefont
  {{Deffner}}}\ and\ \bibinfo {author} {\bibfnamefont {E.}~\bibnamefont
  {{Lutz}}},\ }\href {\doibase 10.1103/PhysRevLett.111.010402} {\bibfield
  {journal} {\bibinfo  {journal} {Physical Review Letters}\ }\textbf {\bibinfo
  {volume} {111}},\ \bibinfo {eid} {010402} (\bibinfo {year}
  {2013})}\BibitemShut {NoStop}%
\bibitem [{\citenamefont {Marvian}\ \emph {et~al.}(2016)\citenamefont
  {Marvian}, \citenamefont {Spekkens},\ and\ \citenamefont
  {Zanardi}}]{MarvianPRA2016}%
  \BibitemOpen
  \bibfield  {author} {\bibinfo {author} {\bibfnamefont {I.}~\bibnamefont
  {Marvian}}, \bibinfo {author} {\bibfnamefont {R.~W.}\ \bibnamefont
  {Spekkens}}, \ and\ \bibinfo {author} {\bibfnamefont {P.}~\bibnamefont
  {Zanardi}},\ }\href {\doibase 10.1103/PhysRevA.93.052331} {\bibfield
  {journal} {\bibinfo  {journal} {Phys. Rev. A}\ }\textbf {\bibinfo {volume}
  {93}},\ \bibinfo {pages} {052331} (\bibinfo {year} {2016})}\BibitemShut
  {NoStop}%
\bibitem [{\citenamefont {Shanahan}\ \emph {et~al.}(2018)\citenamefont
  {Shanahan}, \citenamefont {Chenu}, \citenamefont {Margolus},\ and\
  \citenamefont {del Campo}}]{Shanahan18}%
  \BibitemOpen
  \bibfield  {author} {\bibinfo {author} {\bibfnamefont {B.}~\bibnamefont
  {Shanahan}}, \bibinfo {author} {\bibfnamefont {A.}~\bibnamefont {Chenu}},
  \bibinfo {author} {\bibfnamefont {N.}~\bibnamefont {Margolus}}, \ and\
  \bibinfo {author} {\bibfnamefont {A.}~\bibnamefont {del Campo}},\ }\href
  {\doibase 10.1103/PhysRevLett.120.070401} {\bibfield  {journal} {\bibinfo
  {journal} {Phys. Rev. Lett.}\ }\textbf {\bibinfo {volume} {120}},\ \bibinfo
  {pages} {070401} (\bibinfo {year} {2018})}\BibitemShut {NoStop}%
\bibitem [{\citenamefont {Deffner}\ and\ \citenamefont
  {Campbell}(2017)}]{DeffnerCampbell17}%
  \BibitemOpen
  \bibfield  {author} {\bibinfo {author} {\bibfnamefont {S.}~\bibnamefont
  {Deffner}}\ and\ \bibinfo {author} {\bibfnamefont {S.}~\bibnamefont
  {Campbell}},\ }\href {\doibase 10.1088/1751-8121/aa86c6} {\bibfield
  {journal} {\bibinfo  {journal} {Journal of Physics A: Mathematical and
  Theoretical}\ }\textbf {\bibinfo {volume} {50}},\ \bibinfo {pages} {453001}
  (\bibinfo {year} {2017})}\BibitemShut {NoStop}%
\bibitem [{\citenamefont {An}\ \emph {et~al.}(2016)\citenamefont {An},
  \citenamefont {Lv}, \citenamefont {del Campo},\ and\ \citenamefont
  {Kim}}]{An16}%
  \BibitemOpen
  \bibfield  {author} {\bibinfo {author} {\bibfnamefont {S.}~\bibnamefont
  {An}}, \bibinfo {author} {\bibfnamefont {D.}~\bibnamefont {Lv}}, \bibinfo
  {author} {\bibfnamefont {A.}~\bibnamefont {del Campo}}, \ and\ \bibinfo
  {author} {\bibfnamefont {K.}~\bibnamefont {Kim}},\ }\href
  {https://doi.org/10.1038/ncomms12999} {\bibfield  {journal} {\bibinfo
  {journal} {Nature Communications}\ }\textbf {\bibinfo {volume} {7}},\
  \bibinfo {pages} {12999 EP } (\bibinfo {year} {2016})}\BibitemShut {NoStop}%
\bibitem [{\citenamefont {Campbell}\ and\ \citenamefont
  {Deffner}(2017)}]{DeffnerPRL2017}%
  \BibitemOpen
  \bibfield  {author} {\bibinfo {author} {\bibfnamefont {S.}~\bibnamefont
  {Campbell}}\ and\ \bibinfo {author} {\bibfnamefont {S.}~\bibnamefont
  {Deffner}},\ }\href {\doibase 10.1103/PhysRevLett.118.100601} {\bibfield
  {journal} {\bibinfo  {journal} {Phys. Rev. Lett.}\ }\textbf {\bibinfo
  {volume} {118}},\ \bibinfo {pages} {100601} (\bibinfo {year}
  {2017})}\BibitemShut {NoStop}%
\bibitem [{\citenamefont {{Ito}}\ and\ \citenamefont
  {{Miyadera}}(2017)}]{Ito2017arXiv}%
  \BibitemOpen
  \bibfield  {author} {\bibinfo {author} {\bibfnamefont {K.}~\bibnamefont
  {{Ito}}}\ and\ \bibinfo {author} {\bibfnamefont {T.}~\bibnamefont
  {{Miyadera}}},\ }\href@noop {} {\bibfield  {journal} {\bibinfo  {journal}
  {ArXiv e-prints}\ } (\bibinfo {year} {2017})},\ \Eprint
  {http://arxiv.org/abs/1711.02322} {arXiv:1711.02322 [quant-ph]} \BibitemShut
  {NoStop}%
\bibitem [{\citenamefont {Andolina}\ \emph {et~al.}(2018)\citenamefont
  {Andolina}, \citenamefont {Farina}, \citenamefont {Mari}, \citenamefont
  {Pellegrini}, \citenamefont {Giovannetti},\ and\ \citenamefont
  {Polini}}]{PoliniPRB2018}%
  \BibitemOpen
  \bibfield  {author} {\bibinfo {author} {\bibfnamefont {G.~M.}\ \bibnamefont
  {Andolina}}, \bibinfo {author} {\bibfnamefont {D.}~\bibnamefont {Farina}},
  \bibinfo {author} {\bibfnamefont {A.}~\bibnamefont {Mari}}, \bibinfo {author}
  {\bibfnamefont {V.}~\bibnamefont {Pellegrini}}, \bibinfo {author}
  {\bibfnamefont {V.}~\bibnamefont {Giovannetti}}, \ and\ \bibinfo {author}
  {\bibfnamefont {M.}~\bibnamefont {Polini}},\ }\href {\doibase
  10.1103/PhysRevB.98.205423} {\bibfield  {journal} {\bibinfo  {journal} {Phys.
  Rev. B}\ }\textbf {\bibinfo {volume} {98}},\ \bibinfo {pages} {205423}
  (\bibinfo {year} {2018})}\BibitemShut {NoStop}%
\bibitem [{\citenamefont {Juli\`a-Farr\'e}\ \emph
  {et~al.}(2020{\natexlab{a}})\citenamefont {Juli\`a-Farr\'e}, \citenamefont
  {Salamon}, \citenamefont {Riera}, \citenamefont {Bera},\ and\ \citenamefont
  {Lewenstein}}]{LewensteinBatteries18}%
  \BibitemOpen
  \bibfield  {author} {\bibinfo {author} {\bibfnamefont {S.}~\bibnamefont
  {Juli\`a-Farr\'e}}, \bibinfo {author} {\bibfnamefont {T.}~\bibnamefont
  {Salamon}}, \bibinfo {author} {\bibfnamefont {A.}~\bibnamefont {Riera}},
  \bibinfo {author} {\bibfnamefont {M.~N.}\ \bibnamefont {Bera}}, \ and\
  \bibinfo {author} {\bibfnamefont {M.}~\bibnamefont {Lewenstein}},\ }\href
  {\doibase 10.1103/PhysRevResearch.2.023113} {\bibfield  {journal} {\bibinfo
  {journal} {Phys. Rev. Research}\ }\textbf {\bibinfo {volume} {2}},\ \bibinfo
  {pages} {023113} (\bibinfo {year} {2020}{\natexlab{a}})}\BibitemShut
  {NoStop}%
\bibitem [{\citenamefont {Andolina}\ \emph
  {et~al.}(2019{\natexlab{b}})\citenamefont {Andolina}, \citenamefont {Keck},
  \citenamefont {Mari}, \citenamefont {Giovannetti},\ and\ \citenamefont
  {Polini}}]{PoliniPRB2019}%
  \BibitemOpen
  \bibfield  {author} {\bibinfo {author} {\bibfnamefont {G.~M.}\ \bibnamefont
  {Andolina}}, \bibinfo {author} {\bibfnamefont {M.}~\bibnamefont {Keck}},
  \bibinfo {author} {\bibfnamefont {A.}~\bibnamefont {Mari}}, \bibinfo {author}
  {\bibfnamefont {V.}~\bibnamefont {Giovannetti}}, \ and\ \bibinfo {author}
  {\bibfnamefont {M.}~\bibnamefont {Polini}},\ }\href {\doibase
  10.1103/PhysRevB.99.205437} {\bibfield  {journal} {\bibinfo  {journal} {Phys.
  Rev. B}\ }\textbf {\bibinfo {volume} {99}},\ \bibinfo {pages} {205437}
  (\bibinfo {year} {2019}{\natexlab{b}})}\BibitemShut {NoStop}%
\bibitem [{\citenamefont {{Skrzypczyk}}\ \emph
  {et~al.}(2014{\natexlab{b}})\citenamefont {{Skrzypczyk}}, \citenamefont
  {{Short}},\ and\ \citenamefont {{Popescu}}}]{workextractionPopescuNatComm14}%
  \BibitemOpen
  \bibfield  {author} {\bibinfo {author} {\bibfnamefont {P.}~\bibnamefont
  {{Skrzypczyk}}}, \bibinfo {author} {\bibfnamefont {A.~J.}\ \bibnamefont
  {{Short}}}, \ and\ \bibinfo {author} {\bibfnamefont {S.}~\bibnamefont
  {{Popescu}}},\ }\href {\doibase 10.1038/ncomms5185} {\bibfield  {journal}
  {\bibinfo  {journal} {Nature Communications}\ }\textbf {\bibinfo {volume}
  {5}},\ \bibinfo {eid} {4185} (\bibinfo {year}
  {2014}{\natexlab{b}})}\BibitemShut {NoStop}%
\bibitem [{\citenamefont {M\"uller}(2018{\natexlab{b}})}]{MullerPRX2018}%
  \BibitemOpen
  \bibfield  {author} {\bibinfo {author} {\bibfnamefont {M.~P.}\ \bibnamefont
  {M\"uller}},\ }\href {\doibase 10.1103/PhysRevX.8.041051} {\bibfield
  {journal} {\bibinfo  {journal} {Phys. Rev. X}\ }\textbf {\bibinfo {volume}
  {8}},\ \bibinfo {pages} {041051} (\bibinfo {year}
  {2018}{\natexlab{b}})}\BibitemShut {NoStop}%
\bibitem [{\citenamefont {Allahverdyan}\ and\ \citenamefont
  {Nieuwenhuizen}(2005)}]{AllahverdyanWORKOPPRE2005}%
  \BibitemOpen
  \bibfield  {author} {\bibinfo {author} {\bibfnamefont {A.~E.}\ \bibnamefont
  {Allahverdyan}}\ and\ \bibinfo {author} {\bibfnamefont {T.~M.}\ \bibnamefont
  {Nieuwenhuizen}},\ }\href {\doibase 10.1103/PhysRevE.71.066102} {\bibfield
  {journal} {\bibinfo  {journal} {Phys. Rev. E}\ }\textbf {\bibinfo {volume}
  {71}},\ \bibinfo {pages} {066102} (\bibinfo {year} {2005})}\BibitemShut
  {NoStop}%
\bibitem [{\citenamefont {Talkner}\ \emph {et~al.}(2007)\citenamefont
  {Talkner}, \citenamefont {Lutz},\ and\ \citenamefont
  {H\"anggi}}]{TalknerWORKOPPRE2007}%
  \BibitemOpen
  \bibfield  {author} {\bibinfo {author} {\bibfnamefont {P.}~\bibnamefont
  {Talkner}}, \bibinfo {author} {\bibfnamefont {E.}~\bibnamefont {Lutz}}, \
  and\ \bibinfo {author} {\bibfnamefont {P.}~\bibnamefont {H\"anggi}},\ }\href
  {\doibase 10.1103/PhysRevE.75.050102} {\bibfield  {journal} {\bibinfo
  {journal} {Phys. Rev. E}\ }\textbf {\bibinfo {volume} {75}},\ \bibinfo
  {pages} {050102} (\bibinfo {year} {2007})}\BibitemShut {NoStop}%
\bibitem [{\citenamefont {Perarnau-Llobet}\ \emph {et~al.}(2017)\citenamefont
  {Perarnau-Llobet}, \citenamefont {B\"aumer}, \citenamefont {Hovhannisyan},
  \citenamefont {Huber},\ and\ \citenamefont {Acin}}]{MartiWORKPRL2017}%
  \BibitemOpen
  \bibfield  {author} {\bibinfo {author} {\bibfnamefont {M.}~\bibnamefont
  {Perarnau-Llobet}}, \bibinfo {author} {\bibfnamefont {E.}~\bibnamefont
  {B\"aumer}}, \bibinfo {author} {\bibfnamefont {K.~V.}\ \bibnamefont
  {Hovhannisyan}}, \bibinfo {author} {\bibfnamefont {M.}~\bibnamefont {Huber}},
  \ and\ \bibinfo {author} {\bibfnamefont {A.}~\bibnamefont {Acin}},\ }\href
  {\doibase 10.1103/PhysRevLett.118.070601} {\bibfield  {journal} {\bibinfo
  {journal} {Phys. Rev. Lett.}\ }\textbf {\bibinfo {volume} {118}},\ \bibinfo
  {pages} {070601} (\bibinfo {year} {2017})}\BibitemShut {NoStop}%
\bibitem [{\citenamefont {Rodr\'{\i}guez-Rosario}\ \emph
  {et~al.}(2011)\citenamefont {Rodr\'{\i}guez-Rosario}, \citenamefont {Kimura},
  \citenamefont {Imai},\ and\ \citenamefont
  {Aspuru-Guzik}}]{Aspuru-GuzikPRL2011}%
  \BibitemOpen
  \bibfield  {author} {\bibinfo {author} {\bibfnamefont {C.~A.}\ \bibnamefont
  {Rodr\'{\i}guez-Rosario}}, \bibinfo {author} {\bibfnamefont {G.}~\bibnamefont
  {Kimura}}, \bibinfo {author} {\bibfnamefont {H.}~\bibnamefont {Imai}}, \ and\
  \bibinfo {author} {\bibfnamefont {A.}~\bibnamefont {Aspuru-Guzik}},\ }\href
  {\doibase 10.1103/PhysRevLett.106.050403} {\bibfield  {journal} {\bibinfo
  {journal} {Phys. Rev. Lett.}\ }\textbf {\bibinfo {volume} {106}},\ \bibinfo
  {pages} {050403} (\bibinfo {year} {2011})}\BibitemShut {NoStop}%
\bibitem [{\citenamefont {Das}\ \emph {et~al.}(2018)\citenamefont {Das},
  \citenamefont {Khatri}, \citenamefont {Siopsis},\ and\ \citenamefont
  {Wilde}}]{DasJMathPhys2018}%
  \BibitemOpen
  \bibfield  {author} {\bibinfo {author} {\bibfnamefont {S.}~\bibnamefont
  {Das}}, \bibinfo {author} {\bibfnamefont {S.}~\bibnamefont {Khatri}},
  \bibinfo {author} {\bibfnamefont {G.}~\bibnamefont {Siopsis}}, \ and\
  \bibinfo {author} {\bibfnamefont {M.~M.}\ \bibnamefont {Wilde}},\ }\href
  {\doibase 10.1063/1.4997044} {\bibfield  {journal} {\bibinfo  {journal}
  {Journal of Mathematical Physics}\ }\textbf {\bibinfo {volume} {59}},\
  \bibinfo {pages} {012205} (\bibinfo {year} {2018})}\BibitemShut {NoStop}%
\bibitem [{\citenamefont {Breuer}\ and\ \citenamefont
  {Petruccione}(2007)}]{Book-Open}%
  \BibitemOpen
  \bibfield  {author} {\bibinfo {author} {\bibfnamefont {H.~P.}\ \bibnamefont
  {Breuer}}\ and\ \bibinfo {author} {\bibfnamefont {F.}~\bibnamefont
  {Petruccione}},\ }\href@noop {} {\emph {\bibinfo {title} {The Theory of Open
  Quantum Systems}}}\ (\bibinfo  {publisher} {Oxford University Press, New
  York},\ \bibinfo {year} {2007})\BibitemShut {NoStop}%
\bibitem [{SM()}]{SM}%
  \BibitemOpen
  \href@noop {} {}\bibinfo {note} {See Supplemental Material for proof of
  bounds for open systems and details of the illustration on a model for a heat
  engine.}\BibitemShut {Stop}%
\bibitem [{\citenamefont {{Linden}}\ \emph {et~al.}(2010)\citenamefont
  {{Linden}}, \citenamefont {{Popescu}},\ and\ \citenamefont
  {{Skrzypczyk}}}]{Linden2010}%
  \BibitemOpen
  \bibfield  {author} {\bibinfo {author} {\bibfnamefont {N.}~\bibnamefont
  {{Linden}}}, \bibinfo {author} {\bibfnamefont {S.}~\bibnamefont {{Popescu}}},
  \ and\ \bibinfo {author} {\bibfnamefont {P.}~\bibnamefont {{Skrzypczyk}}},\
  }\href@noop {} {\bibfield  {journal} {\bibinfo  {journal} {ArXiv e-prints}\ }
  (\bibinfo {year} {2010})},\ \Eprint {http://arxiv.org/abs/1010.6029}
  {arXiv:1010.6029 [quant-ph]} \BibitemShut {NoStop}%
\bibitem [{\citenamefont {Brunner}\ \emph {et~al.}(2012)\citenamefont
  {Brunner}, \citenamefont {Linden}, \citenamefont {Popescu},\ and\
  \citenamefont {Skrzypczyk}}]{Linden2010PRE}%
  \BibitemOpen
  \bibfield  {author} {\bibinfo {author} {\bibfnamefont {N.}~\bibnamefont
  {Brunner}}, \bibinfo {author} {\bibfnamefont {N.}~\bibnamefont {Linden}},
  \bibinfo {author} {\bibfnamefont {S.}~\bibnamefont {Popescu}}, \ and\
  \bibinfo {author} {\bibfnamefont {P.}~\bibnamefont {Skrzypczyk}},\ }\href
  {\doibase 10.1103/PhysRevE.85.051117} {\bibfield  {journal} {\bibinfo
  {journal} {Phys. Rev. E}\ }\textbf {\bibinfo {volume} {85}},\ \bibinfo
  {pages} {051117} (\bibinfo {year} {2012})}\BibitemShut {NoStop}%
\bibitem [{\citenamefont {Garc\'{\i}a-Pintos}\ \emph
  {et~al.}(2021)\citenamefont {Garc\'{\i}a-Pintos}, \citenamefont {Hamma},\
  and\ \citenamefont {del Campo}}]{LPGPreply}%
  \BibitemOpen
  \bibfield  {author} {\bibinfo {author} {\bibfnamefont {L.~P.}\ \bibnamefont
  {Garc\'{\i}a-Pintos}}, \bibinfo {author} {\bibfnamefont {A.}~\bibnamefont
  {Hamma}}, \ and\ \bibinfo {author} {\bibfnamefont {A.}~\bibnamefont {del
  Campo}},\ }\href {\doibase 10.1103/PhysRevLett.127.028902} {\bibfield
  {journal} {\bibinfo  {journal} {Phys. Rev. Lett.}\ }\textbf {\bibinfo
  {volume} {127}},\ \bibinfo {pages} {028902} (\bibinfo {year}
  {2021})}\BibitemShut {NoStop}%
\bibitem [{\citenamefont {Garc\'{\i}a-Pintos}\ \emph
  {et~al.}(2020)\citenamefont {Garc\'{\i}a-Pintos}, \citenamefont {Hamma},\
  and\ \citenamefont {del Campo}}]{garcapintos2019fluctuations}%
  \BibitemOpen
  \bibfield  {author} {\bibinfo {author} {\bibfnamefont {L.~P.}\ \bibnamefont
  {Garc\'{\i}a-Pintos}}, \bibinfo {author} {\bibfnamefont {A.}~\bibnamefont
  {Hamma}}, \ and\ \bibinfo {author} {\bibfnamefont {A.}~\bibnamefont {del
  Campo}},\ }\href {\doibase 10.1103/PhysRevLett.125.040601} {\bibfield
  {journal} {\bibinfo  {journal} {Phys. Rev. Lett.}\ }\textbf {\bibinfo
  {volume} {125}},\ \bibinfo {pages} {040601} (\bibinfo {year}
  {2020})}\BibitemShut {NoStop}%
\bibitem [{\citenamefont {Cusumano}\ and\ \citenamefont
  {Rudnicki}(2021)}]{cusumano2021comment}%
  \BibitemOpen
  \bibfield  {author} {\bibinfo {author} {\bibfnamefont {S.}~\bibnamefont
  {Cusumano}}\ and\ \bibinfo {author} {\bibfnamefont {L.}~\bibnamefont
  {Rudnicki}},\ }\href {\doibase 10.1103/PhysRevLett.127.028901} {\bibfield
  {journal} {\bibinfo  {journal} {Phys. Rev. Lett.}\ }\textbf {\bibinfo
  {volume} {127}},\ \bibinfo {pages} {028901} (\bibinfo {year}
  {2021})}\BibitemShut {NoStop}%
\bibitem [{\citenamefont {Helstrom}(1969)}]{helstrom1969quantum}%
  \BibitemOpen
  \bibfield  {author} {\bibinfo {author} {\bibfnamefont {C.~W.}\ \bibnamefont
  {Helstrom}},\ }\href {\doibase 10.1007/BF01007479} {\emph {\bibinfo {title}
  {Quantum detection and estimation theory}}},\ Vol.~\bibinfo {volume} {1}\
  (\bibinfo  {publisher} {Springer},\ \bibinfo {year} {1969})\ pp.\ \bibinfo
  {pages} {231--252}\BibitemShut {NoStop}%
\bibitem [{\citenamefont {Holevo}(2011)}]{holevo2011probabilistic}%
  \BibitemOpen
  \bibfield  {author} {\bibinfo {author} {\bibfnamefont {A.~S.}\ \bibnamefont
  {Holevo}},\ }\href {\doibase 10.1007/978-88-7642-378-9} {\emph {\bibinfo
  {title} {Probabilistic and statistical aspects of quantum theory}}},\
  Vol.~\bibinfo {volume} {1}\ (\bibinfo  {publisher} {Springer Science \&
  Business Media},\ \bibinfo {year} {2011})\BibitemShut {NoStop}%
\bibitem [{\citenamefont {Braunstein}\ and\ \citenamefont
  {Caves}(1994)}]{BraunsteinCaves1994}%
  \BibitemOpen
  \bibfield  {author} {\bibinfo {author} {\bibfnamefont {S.~L.}\ \bibnamefont
  {Braunstein}}\ and\ \bibinfo {author} {\bibfnamefont {C.~M.}\ \bibnamefont
  {Caves}},\ }\href {\doibase 10.1103/PhysRevLett.72.3439} {\bibfield
  {journal} {\bibinfo  {journal} {Phys. Rev. Lett.}\ }\textbf {\bibinfo
  {volume} {72}},\ \bibinfo {pages} {3439} (\bibinfo {year}
  {1994})}\BibitemShut {NoStop}%
\bibitem [{\citenamefont {Parisi}(2009)}]{paris2009quantum}%
  \BibitemOpen
  \bibfield  {author} {\bibinfo {author} {\bibfnamefont {M.~G.}\ \bibnamefont
  {Parisi}},\ }\href {\doibase 10.1142/S0219749909004839} {\bibfield  {journal}
  {\bibinfo  {journal} {International Journal of Quantum Information}\ }\textbf
  {\bibinfo {volume} {7}},\ \bibinfo {pages} {125} (\bibinfo {year}
  {2009})}\BibitemShut {NoStop}%
\bibitem [{\citenamefont {Sidhu}\ and\ \citenamefont
  {Kok}(2020)}]{sidhu_geometric_2020}%
  \BibitemOpen
  \bibfield  {author} {\bibinfo {author} {\bibfnamefont {J.~S.}\ \bibnamefont
  {Sidhu}}\ and\ \bibinfo {author} {\bibfnamefont {P.}~\bibnamefont {Kok}},\
  }\href {\doibase 10.1116/1.5119961} {\bibfield  {journal} {\bibinfo
  {journal} {AVS Quantum Science}\ }\textbf {\bibinfo {volume} {2}},\ \bibinfo
  {pages} {014701} (\bibinfo {year} {2020})}\BibitemShut {NoStop}%
\bibitem [{\citenamefont {Juli\`a-Farr\'e}\ \emph
  {et~al.}(2020{\natexlab{b}})\citenamefont {Juli\`a-Farr\'e}, \citenamefont
  {Salamon}, \citenamefont {Riera}, \citenamefont {Bera},\ and\ \citenamefont
  {Lewenstein}}]{PhysRevResearch.2.023113}%
  \BibitemOpen
  \bibfield  {author} {\bibinfo {author} {\bibfnamefont {S.}~\bibnamefont
  {Juli\`a-Farr\'e}}, \bibinfo {author} {\bibfnamefont {T.}~\bibnamefont
  {Salamon}}, \bibinfo {author} {\bibfnamefont {A.}~\bibnamefont {Riera}},
  \bibinfo {author} {\bibfnamefont {M.~N.}\ \bibnamefont {Bera}}, \ and\
  \bibinfo {author} {\bibfnamefont {M.}~\bibnamefont {Lewenstein}},\ }\href
  {\doibase 10.1103/PhysRevResearch.2.023113} {\bibfield  {journal} {\bibinfo
  {journal} {Phys. Rev. Research}\ }\textbf {\bibinfo {volume} {2}},\ \bibinfo
  {pages} {023113} (\bibinfo {year} {2020}{\natexlab{b}})}\BibitemShut
  {NoStop}%
\end{thebibliography}%

%\widetext
\clearpage
\appendix

\section*{Garc\'ia-Pintos et. al. Reply~\cite{LPGPreply}}

\emph{
We acknowledge that a derivation reported in~\cite{garcapintos2019fluctuations} is incorrect, as pointed out by Cusumano and Rudnicki~\cite{cusumano2021comment}. We respond by giving a correct proof of the claim ``fluctuations in the free energy operator upper bound the charging power of a quantum battery'' that we made in the paper.
}

 \ 

We thank Cusumano and Rudnicki for bringing to our attention the mistakes made in~\cite{garcapintos2019fluctuations}. Indeed, the conclusion that follows Eq.~(17), and Eq.~(18), are incorrect. As they have noted in~\cite{cusumano2021comment}, without further conditions open quantum systems can have a non-zero charging power for batteries in an eigenstate of $\work$.

Even though expression~(18) is incorrect, 
we stress that it holds that ``fluctuations in the free energy operator bound the charging power of a quantum battery'', as claimed in the Letter. Equation~(12) showed this for isolated quantum batteries, and this holds for open quantum batteries as well.
We present a correct proof of this statement next.

The rate of change of the extractable work satisfies
\begin{align}
P(t) &= \frac{d \langle \work \rangle_\bat}{dt} \nonumber \\
&= \frac{d}{dt}\tr{H_\bat \rho_\bat} + \beta^{-1} \frac{d}{dt}\tr{\rho_\bat \log(\rho_\bat)} \nonumber \\
&=   \tr{H_\bat \frac{d\rho_\bat}{dt}} + \beta^{-1}  \tr{\frac{d\rho_\bat}{dt} \log(\rho_\bat)} \nonumber \\
&= \tr{\frac{d\rho_\bat}{dt} \work},
\end{align}
where we used that the self-Hamiltonian $H_\bat$ of the battery is time-independent~\cite{garcapintos2019fluctuations} and that $\frac{d}{dt}\tr{\rho_\bat \log(\rho_\bat)} \nobreak = \nobreak \tr{\frac{d\rho_\bat}{dt} \log(\rho_\bat)}$ \LP{holds for finite dimensional systems as well as for states $\rho_\bat$ with constant kernel~\cite{DasJMathPhys2018}}.

Defining $\delta \work \nobreak\coloneqq \nobreak \work \nobreak-\nobreak \langle \work \rangle_\bat$ and expressing the state of the battery in its instantaneous eigenbasis $\{\ket{\alpha}\}$ as $\rho_\bat \nobreak=\nobreak \sum_\alpha p_\alpha \ket{\alpha}\!\bra{\alpha}$, the Cauchy-Schwarz inequality yields 
\begin{align}
\label{eq:generalderivation}
\left| P(t) \right| &= \left| \tr{\frac{d\rho_\bat}{dt} \delta \work} \right| = \bigg| \sum_{{\alpha \beta}} \delta \work_{\alpha \beta} \bra{\beta} \tfrac{d\rho_\bat}{dt} \ket{\alpha} \bigg| \nonumber \\ 
&= \bigg| \sum_{\alpha \beta}  \tfrac{\sqrt{p_\alpha + p_\beta}}{\sqrt{p_\alpha + p_\beta}} \delta \work_{\alpha \beta} \bra{\beta} \tfrac{d\rho_\bat}{dt} \ket{\alpha}  \bigg| \nonumber \\
&\leq \sqrt{ \frac{1}{2} \sum_{\alpha \beta} (p_\alpha + p_\beta) |\delta \work_{\alpha \beta}|^2} \sqrt{ 2 \sum_{\alpha \beta} \tfrac{\left| \bra{\beta} \tfrac{d\rho_\bat}{dt} \ket{\alpha}\right|^2}{p_\alpha + p_\beta}},
\end{align}
where we used that $\tfrac{d\rho_\bat}{dt}$ is traceless and assumed differentiable dynamics.

The first factor in the bound is related to the fluctuations in the free energy operator, with $(\sigma_\work )^2 \nobreak = \nobreak \tr{\rho_\bat (\delta \work)^2} \nobreak = \nobreak \sum_{\alpha \beta} p_\alpha  |\delta \work_{\alpha \beta}|^2 $.
The second factor is related to the quantum Fisher information parametrized by time~\cite{helstrom1969quantum,
 holevo2011probabilistic,
 BraunsteinCaves1994},
\begin{align}
\label{eq:qFisher}
\info \coloneqq 2 \sum_{\alpha \beta}  \frac{\left| \bra{\beta} \frac{d \rho_\bat}{d t} \ket{\alpha}\right|^2}{p_\alpha + p_\beta},
\end{align}
which can diverge in instances of time when levels with $p_\alpha = 0$ begin populating, but is otherwise finite for states with constant kernel.
The Fisher information $\info$ characterizes the sensitivity of the state to time-translations, and determines the ultimate precision with which time can be estimated in the system
(see~\cite{paris2009quantum,sidhu_geometric_2020} for accessible interpretations of the quantum Fisher information and its role in the study of quantum parameter estimation).
The Fisher information has also been shown to bound the rate of change of the energy of quantum batteries~\cite{PhysRevResearch.2.023113}, but note how the proof in Eq.~\eqref{eq:generalderivation}, in contrast, also incorporates the role of entropy to the extractable work.

We have thus proven that an interplay between the fluctuations in the free energy operator and the quantum Fisher information bound the rate of change of the extractable work,
\begin{align}
\label{eq:generalbound}
\left| P(t) \right| \leq \sigma_\work \, \sqrt{\info}.
\end{align}
Given two states with identical Fisher information $\info$, the one with higher fluctuations in free energy can sustain a higher charging power.

Care needs to be taken for states whose rank is changing, since in those cases unpopulated states with $ p_\alpha = 0 $ contribute to a divergent Fisher information.
This is what happens for the initial state $\rho_\bat(t_0) = \ket{k_0}\!\bra{k_0}$ analyzed in~\cite{garcapintos2019fluctuations} and by Cusumano and Rudnicki, since at $t_0$ a pure state is evolving to a mixed one. 
On the contrary, if at any $t \geq t_0$ the kernel of the state of the system is not changing, for instance for states with full support, Eq.~\eqref{eq:generalbound} bounds power.

\clearpage

\section*{Appendix}

\subsection{Limits to work extraction -- open systems}

The formalism introduced for isolated systems allows to extend the analysis to include an open-system description of the dynamics of the battery. 
Using that $ d\rho /dt = -i [H,\rho ]$, the state of the battery evolves according to 
\begin{align}
\label{eq-app:openevol}
\frac{d}{dt} \rho_\bat &= -i \trs{\sys \bath \anc}{\left [H,\rho  \right]} \nonumber \\
&= -i \trs{\sys \bath \anc}{\left [ \id_{\sys \bath \anc} \otimes H_\bat + V,\rho \right]} \nonumber \\
& = -i[H_\bat,\rho_\bat ] -i \trs{\sys \bath \anc}{\left [ V,\rho \right]}.
\end{align}
In the Markovian limit the evolution is well approximated by 
\begin{align}
\frac{d}{dt} \rho_\bat  &\approx -i[H_\bat,\rho_\bat ] -i[\tilde H_\bat,\rho_\bat ]  \\
&+ \sum_j \gamma_j \left( L_j \rho_\bat  L_j^\dag - \frac{1}{2} \left\{ L_j^\dag L_j ,\rho_\bat \right\} \right) ,\nonumber
\end{align}
where $\tilde H_\bat$ accounts for the unitary part of the evolution due to the interactions, 
the Lindblad operators $L_j$ characterize the effect of the interaction of the battery with the remaining systems, and
the rates $\gamma_j$ are non-negative~\cite{Book-Open}.
Then,
\begin{align}
& \trs{\sys \bath \anc}{\left [ V,\rho \right]} \approx [\tilde H_\bat,\rho_\bat ]  \\
 &\quad \quad \quad  + i \sum_j \gamma_j \left( L_j \rho_\bat L_j^\dag - \frac{1}{2} \left\{ L_j^\dag L_j ,\rho_\bat \right\} \right). \nonumber
\end{align}
This allows to derive an alternative bound on the charging power that only depends on the state of the battery and the Lindblad operators.
Defining $\delta \mathcal{F} \equiv \mathcal{F} -\langle \mathcal{F} \rangle_\bat$, and using that 
\begin{align}
P_t &= -i \tr{  \rho \left[\mathcal{F} \otimes \id_{\sys \bath \anc}, V \right]} \label{eq-app:coherenceBattery},
\end{align}
it holds that
\begin{align}
\label{eq-app:derivationOpen}
| P_t | &= \Big| \tr{\rho [\work \otimes \id_{\sys \bath \anc} ,V ]} \Big|   = \Big| \tr{\delta \mathcal{F} \otimes \id_{\sys \bath \anc}   \left[ V ,\rho \right]} \Big|  \nonumber \\
&  = \Big| \trs{\bat}{\delta \mathcal{F} \trs{\sys \bath \anc}{  \left[ V ,\rho \right] }} \Big|  \nonumber \\
&=   \bigg| \trs{\bat}{\delta \mathcal{F} [\tilde H_\bat,\rho_\bat ] }  \nonumber \\
&\!\!\!\!\!\!\!  + i \sum_j  \gamma_j     \trs{\bat}{\delta \mathcal{F} \left(  L_j \rho_\bat L_j^\dag - \frac{1}{2} \left\{ L_j^\dag L_j ,\rho_\bat \right\}   \right) } \bigg| \nonumber \\
&\le   \Big| \trs{\bat}{\delta \mathcal{F} [\tilde H_\bat,\rho_\bat ] }  \Big| \nonumber \\
&\!\!\!\!\!\!\!  + \sum_j  \gamma_j   \left| \trs{\bat}{\delta \mathcal{F} \left(  L_j \rho_\bat  L_j^\dag - \frac{1}{2} \left\{ L_j^\dag L_j ,\rho_\bat \right\}   \right) } \right| \nonumber \\
& =  \Big| \trs{\bat}{\delta \mathcal{F} [\tilde H_\bat,\rho_\bat ] }  \Big| \nonumber \\
&\!\!\!\!\!\!\!  + \sum_j  \frac{\gamma_j}{2}  \left| \trs{\bat}{  [ L_j^\dag,\delta \work ] L_j \rho_\bat + L_j^\dag[ \delta \work ,L_j ] \rho_\bat } \right| 
\end{align}
The Cauchy-Schwarz inequality implies that
\begin{align}
&\left| \trs{\bat}{  L_j^\dag [\delta \work , L_j ] \rho_\bat } \right|^2 \nonumber \\
&\qquad \leq \trs{\bat}{ \big| [ \delta \work, L_j ] \big|^2 \rho_\bat } \trs{\bat}{ \big| L_j^\dag \big|^2 \rho_\bat } \nonumber \\
&\qquad \leq \trs{\bat}{ \big| [ \delta \work, L_j ] \big|^2 \rho_\bat }  \| L_j \|^2 \nonumber \\
&\qquad \leq \left\langle \big| [ \delta \work, L_j ] \big|^2 \right\rangle  \| L_j \|^2,
\end{align} 
and
\begin{align}
 \left| \trs{\bat}{\delta \mathcal{F} [\tilde H_\bat,\rho_\bat ] }  \right|^2 \leq 4 \sigma^2_{\work}(t) \sigma^2_{\tilde H_\bat}(t).
\end{align}
Then, 
\begin{align}
\label{eq-app:boundopen}
| P_t | &\leq  2 \sigma_{\work}(t) \sigma_{\tilde H_\bat}(t) + \sum_j   \gamma_j  \sqrt{\left\langle \big| [ \delta \work, L_j ] \big|^2 \right\rangle} \|L_j \|.
\end{align}
Interestingly, for the open-system case it also holds that the charging power is null unless there exist fluctuations in the \LP{extractable work of} the battery. In order to see this, let $\delta \work = \sum_j w_j \ket{j} \bra{j}$. In the eigenbasis of $\delta \work$, one can write
\begin{align}
\left\langle \big| [ \delta \work, L_j ] \big|^2 \right\rangle = \sum_{jkl} \rho_{jk} L_{kl} L^\dag_{lj} \left( w_l^2 - w_l w_j - w_l w_k + w_j w_k \right).
\end{align}
Thus, for states $\rho_\bat = \ket{j} \bra{j}$ with a deterministic amount of \LP{free energy characterized by} $w_j$, it follows that $\left\langle \big| [ \delta \work, L_j ] \big|^2 \right\rangle = 0$ and that $\sigma_\work = 0$, leading to
\begin{align}
|P_t| = 0.
\end{align}
In contrast, states with support on many eigenstates of $\work$ can sustain a higher charging power, as $\left\langle \big| [ \delta \work, L_j ] \big|^2 \right\rangle \neq 0$ and $\sigma_\work \neq 0$.

Note, too, that the third line of Eq.~\eqref{eq-app:derivationOpen} suggests that generating coherence in the eigenbasis of the \LP{free energy} operator $\work$, or of the shifted Hamiltionian $\tilde H_\bat$, may be beneficial for the charging power output.

Equation~\eqref{eq-app:boundopen} can be further bounded for hermitian Lindblad operators, corresponding to pure dephasing. In that case, applying the Cauchy-Schwarz inequality, and using that $|\tr{ABC}| = |\tr{ACB}|$ for hermitian operators, gives
\begin{align}
\left\langle \big| [ \delta \work, L_j ] \big|^2 \right\rangle &\leq 4 \| L_j\|^2 \trs{\bat}{\rho_\bat (\delta \work)^2} \nonumber \\
&= 4 \| L_j\|^2 \sigma_\work^2(t).
\end{align}
That is, for hermitian Lindblad operators the \LP{charging} rate is bounded by
\begin{align}
\label{eq-app:boundopenHermi}
| P_t | &\leq  \sigma_{\work}(t) \left( 2 \sigma_{\tilde H_\bat}(t) + \sum_j   2 \gamma_j  \|L_j \|^2 \right).
\end{align}

\subsection{Bounds on charging rate -- exact solution at zero temperature}

For the case of a reference bath at zero temperature, 
the rate at which \LP{the extractable work --in this case energy--} changes in the battery is 
\begin{align}
P(t) = \frac{d \tr{\rho_\bat  H_\bat}}{dt}.
\end{align}

Denoting the standard deviations of the \LP{extractable} work and of the battery interaction Hamiltonian by $\sigma_\work$ and $\sigma_V$ respectively, 
\begin{align}
\sigma_\work^2(t) &= \tr{ \rho  H_\bat^2 } - \tr{ \rho  H_\bat}^2 \\
% \tr{ \rho_t \left(H_\bat - U_\bat \right)^2 } \\
\sigma_V^2(t) &= \tr{ \rho V^2} - \tr{ \rho V}^2 ,
%\tr{ \rho_t \left(V - v \right)^2 },
\end{align}
%eq.~\eqref{eq:energyboundderivation} becomes
the rate of \LP{extractable} work with reference to a zero-temperature bath satisfies
\begin{align}
%\label{eq:powerbound}
\left|P(t)\right| \le 2 \sigma_{\work}(t) \sigma_V(t).
\end{align}
This implies a trade-off between \LP{extractable }work (energy in this case) charging power and the fluctuations of the \LP{extractable work}: for a fixed interaction with the battery, a desired power input necessarily comes with fluctuations \LP{in extractable work}.

\subsubsection{Toy model of a heat engine}

\begin{figure}[!h]
  \centering
       \hspace{0pt} 
       \includegraphics[scale=.25]{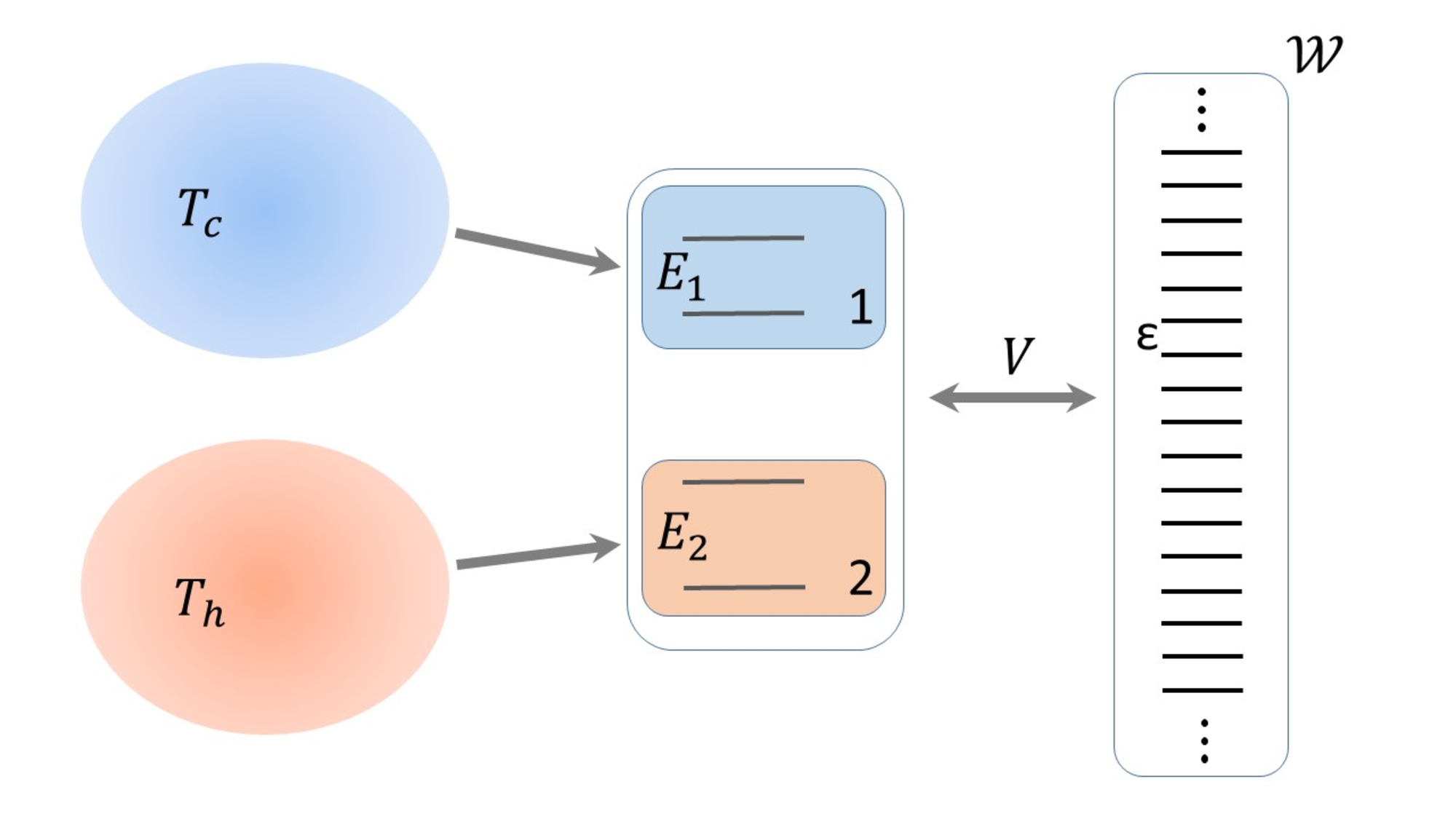}       \caption{\label{fig:engine}
The heat engine stores work in the battery $\bat$ by exploiting heat flowing from the hot to the cold baths. The baths and battery are coupled indirectly via qubits $1$ and $2$, working as a switch in the engine. As heat flows via the switch, the energy and extractable work stored in the battery increase.
      }
\end{figure}

We illustrate the connection between charging power, fluctuations in 
the free energy operator $\work$,
%the stored work,
 and quantum coherence, in a the simple toy model for a heat engine considered in~\cite{Linden2010,Linden2010PRE}. The engine, depicted in fig.~\ref{fig:engine}, consists of a hot bath at temperature $T_h$ and a cold bath at temperature $T_c$ as resources.
Heat flow from the hot to the cold bath is exploited to extract work and store it in the battery, which consists of a Harmonic oscillator unbounded from below, with an energy gap $\epsilon$ in a Hamiltonian
\begin{align}
H_\bat = \sum_{n = -\infty}^\infty n\epsilon \ket{n}_w\!\bra{n}.
\end{align}
The storage device is indirectly coupled to the heat baths via a `switch', consisting of two qubits. Qubit $1$, with energy gap $E_1$, is coupled to the cold bath, while qubit $2$, with energy gap $E_2$, is coupled to the hot bath. 
The qubits have a free Hamiltonian given by
\begin{align}
H_\sys = E_1 \ket{1}_1\!\bra{1} + E_2 \ket{1}_2\!\bra{1},
\end{align}
with energies taken such that $E_2-E_1 = \epsilon$.
The qubits interact with the battery via
\begin{align}
V = g \sum_{n=-\infty}^\infty \Big( \ket{01,n}\!\bra{10,n+1} + \ket{10,n+1}\!\bra{01,n} \Big),
\end{align}
where $g$ is a coupling constant.

In the model, the qubits are assumed to thermalize due to the interaction with the thermal baths, which is modeled by the master equation
\begin{align}
\frac{d\rho}{dt} = -i[H_\sys + H_\bat + V,\rho] + \sum_{i=1}^2 p_i (\tau_i \trs{i}{\rho} - \rho),
\end{align}
where $\rho$ is the density matrix of the storage device and the qubits 1 and 2, and $p_i$ characterizes the rate at which qubit $`i'$ thermalizes due to the interaction with its corresponding bath (see~\cite{Linden2010,Linden2010PRE} for more details of the model).  
Thermal states of the qubits are denoted by
\begin{align}
\tau_i &= r_i\ket{0}_i\!\bra{0} + \overline{r_i} \ket{1}_i\!\bra{1},
\end{align}
with
\begin{align}
\overline{r_1} &= r_1 e^{-\frac{E_1}{kT_c}}, \qquad \overline{r_2} = r_2 e^{-\frac{E_2}{kT_h}}, 
\end{align}
where $k$ is Boltzmann's constant. If $E_1 / T_c > E_2 / T_h$ the device favors net energy exchange from hot to cold bath, working as a heat engine.

In~\cite{Linden2010,Linden2010PRE} the authors show that the work deposited in the battery can be obtained from solving for $e_\bat \equiv \tr{H_\bat \rho_\bat}$,
%We are interested in the work deposited in the battery, along with its statistical properties. 
%The change in energy can be found from solving the equation
\begin{equation}
\frac{d e_\bat}{dt} = \frac{d}{dt}\tr{H_\bat \rho_\bat} = -ig\epsilon \Delta(t), 
\end{equation}
with
\begin{subequations}
\begin{align}
\Delta(t) &= \sum_n \Big( \bra{01,n}\rho\ket{10,n+1} - \bra{10,n+1}\rho\ket{01,n} \Big) \\
\Gamma_1(t) &= \sum_n \Big( \bra{00,n}\rho\ket{00,n} + \bra{01,n}\rho\ket{01,n} \Big) \\
\Gamma_2(t) &= \sum_n \Big( \bra{00,n}\rho\ket{00,n} + \bra{10,n}\rho\ket{10,n} \Big),
\end{align}
\end{subequations}
which obey the set of differential equations
\begin{subequations}
\begin{align}
\frac{d}{dt} \Delta(t) &= 2ig\Big(\Gamma_1(t) - \Gamma_2(t) \Big) - \big( p_1 + p_2 \big) \Delta(t) \\
\frac{d}{dt} \Gamma_1(t) &= ig \Delta(t) + p_1 \Big(r_1 - \Gamma_1(t) \Big) \\
\frac{d}{dt} \Gamma_2(t) &= -ig \Delta(t) + p_2 \Big(r_2 - \Gamma_2(t) \Big).
\end{align}
\end{subequations}

While~\cite{Linden2010,Linden2010PRE} focuses on the general, asymptotic, behavior of the charging power, we are interested in the charging power when initial states with uncertain amounts of \LP{free energy}, and quantum coherence, are taken.
Consider first a state in which qubits $1$ and $2$ are initially in thermal equilibrium with their respective thermal baths, and the battery in an eigenstate of its Hamiltonian, 
\begin{align}
\rho_{\textnormal{diag}} &= \Big( r_1 \ket{0}_1 \bra{0} + \overline{r_1} \ket{1}_1 \bra{1}  \Big)  \nonumber \\
&\Big( r_2 \ket{0}_2 \bra{0} + \overline{r_2} \ket{2}_1 \bra{2}  \Big) \ket{0}_\bat\bra{0}.
\end{align}
Naively, this sounds like an ideal initial state for the battery, with a well-defined deterministic initial amount of \LP{extractable work and energy}, without fluctuations. 
However, for such a state, diagonal in both interaction and battery Hamiltonians, Eq.~(8) %\eqref{eq:coherenceInteraction} and~
%\eqref{eq:coherenceBattery} 
in the main text
implies that the engine initially functions with null power.
Importantly, that the is the case  for incoherent mixture states without deterministic initial \LP{free energy}, e.g. thermal states. 

In order to have non-zero charging power a coherent superposition in the battery and interaction Hamiltonians is needed (note, though, that this may not be sufficient).
Consequently, we consider both the qubits  and battery in a pure state, the latter in a superposition between $N$ energy levels, with equal weights for simplicity:
\begin{align}
\ket{\Psi_N} &= \Big( \sqrt{r_1}\ket{0}_1 + e^{i \theta} \sqrt{\overline{r_1}}\ket{1}_1  \Big) \nonumber \\
&\Big( \sqrt{r_2}\ket{0}_2 + \sqrt{\overline{r_2}}\ket{1}_2  \Big) \frac{1}{\sqrt{N}} \sum_{n=0}^{N-1}\ket{n}_\bat .
\end{align}
The phase $e^{i \theta}$ can change the output of the engine, and even turn the device into a refrigerator during a transient time, for some values of $\theta$.% [[[ilustrarlo en figs?]]].
%\begin{align}
%\ket{\Psi_2} &= \Big( \sqrt{r_1}\ket{0}_1 + e^{i \theta} \sqrt{\overline{r_1}}\ket{1}_1  \Big) \nonumber \\
%&\Big( \sqrt{r_2}\ket{0}_2 + \sqrt{\overline{r_2}}\ket{1}_2  \Big) \frac{1}{\sqrt{2}}\Big( \ket{0}_\bat + \ket{1}_\bat \Big).
%\end{align}

\begin{figure}[!h]
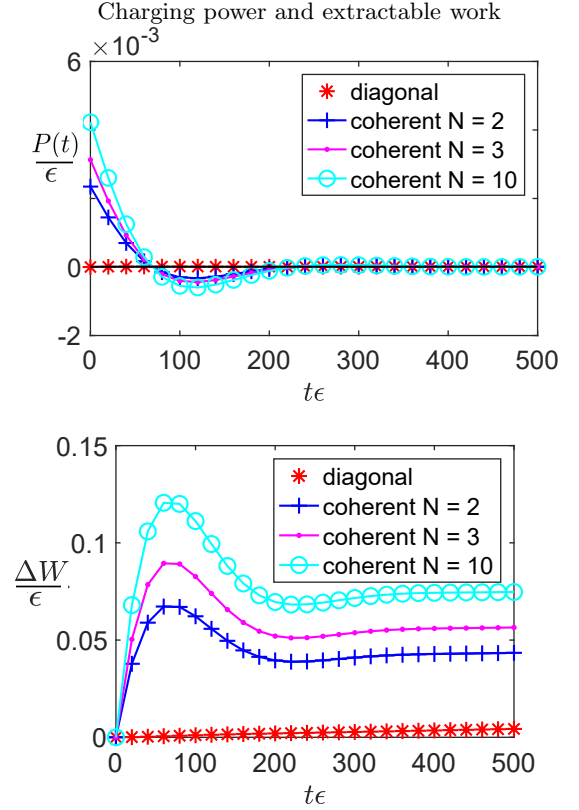

  \centering
       {Charging power and \LP{extractable} work
       \hspace{0 cm}}   \hspace{0pt} 
       \includegraphics[scale=.49]{fig2-power.eps} \hspace{-0.0cm}
       \\
       \vspace{12pt}
\hspace{-22pt}       \includegraphics[scale=.50]{fig2-storedwork.eps} 
  \caption{\label{fig-app:powerandwork}
      Charging power $P(t)$ (top) and change in \LP{extractable} work $\Delta W(t) \equiv W_\textnormal{max}(t) - W_\textnormal{max}(0)$ (bottom) as a function of time for different initial states. The parameters in the model are takes as $p_1 = p_2 = g/\epsilon = 0.01$, and $E_1/T_c = 0.51$ and $E_2/T_h = 0.50$. The phase in the initial states $\ket{\Psi_N}$ is taken $\theta = \pi/2$ to ensure a positive initial charging power (interestingly, for other phases the engine \LP{functions} as a refrigerator).
The incoherent state $\rho_\textnormal{diag}$ (red asterisk) has null charging power initially, as needs be from bound Eq.~(12) in the main text. 
%~\eqref{eq:powerbound}. 
This charging power is slowly increased as the state supports higher fluctuations \LP{in the extractable work}.
A significant increase in the charging power is seen for a coherent superposition between 2 eigenvalues of $H_\bat$. For intermediate times this initial increase is penalized with a regime in which the battery is being discharged though. Nevertheless, there is a significant net increase in the \LP{extractable} work with respect to an incoherent initial state. 
Taking coherent superpositions between more than two states gives an even further increase in charging power and \LP{extractable} work, but the increase is marginal for large superpositions, much smaller than the advantage of going from incoherent to minimal coherent superpositions. 
}
\end{figure}

With such coherent superposition as the initial state one has $\sigma_\work \ge 0$ and $\sigma_v \ge 0$, such that inequality~(8) in the main text %\eqref{eq:coherenceInteraction} and~
%\eqref{eq:coherenceBattery}
 allows to have a non-zero charging power. This is to be compared with the null charging power of state $\rho_\textnormal{diag}$. 
Figure~\ref{fig-app:powerandwork} compares the charging power $P(t)$ and the change in the \LP{extractable} work $\Delta W(t) \equiv W_\textnormal{max}(t) - W_\textnormal{max}(0)$ as a function of time, for initial states given by $\rho_\textnormal{diag}$ and $\ket{\Psi_N}\!\bra{\Psi_N}$, for different values of $N$. 
A superposition between two $H_\bat$ eigenstates indeed causes an increase in the charging power, with respect to the null charging power of the incoherent state. 
Taking superpositions between more levels results in an even higher power.
This is reflected in the total \LP{extractable work of} battery as well, with a considerable increase for coherent superpositions, with the most noticeable advantage achieved when going from incoherent state to $\ket{\Psi_2}\!\bra{\Psi_2}$.

While Fig.~\ref{fig-app:powerandwork} illustrates the higher amount of \LP{extractable} work obtained from states with coherence and non-zero \LP{extractable work} fluctuations, 
it also raises the question of what happens with such fluctuations in \LP{extractable work}. 
%Clearly the mean energy of the battery cannot be the single figure of merit for the usefulness of work. 
In order to study the evolution of the \LP{free energy} variance, we consider
\begin{align}
A(t) \equiv \tr{H_\bat^2\rho_\bat}.
\end{align}
Then
\begin{align}
\sigma_\work^2(t) = A(t) - e_\bat^2(t).
\end{align}
Similar calculations as in~\cite{Linden2010,Linden2010PRE} show that 
\begin{align}
\frac{d}{dt} A(t) = -ig \epsilon^2 \Delta(t) - 2ig \epsilon^2 \alpha(t),
\end{align}
where the system of differential equations is completed with
\begin{subequations}
\begin{align}
\frac{d}{dt} \alpha(t) &= 2ig \Big( \beta(t) - \gamma(t) -\delta(t) \Big) - (p_1+p_2)\alpha(t) \\
\frac{d}{dt} \beta(t) &= ig \alpha(t) -p_1\beta(t) + \frac{p_1 r_1}{\epsilon} e_\bat(t)\\
\frac{d}{dt} \gamma(t) &= -ig \alpha(t) -p_2\gamma(t) + \frac{p_2 r_2}{\epsilon} e_\bat(t) - p_2 r_2\\
\frac{d}{dt} \delta(t) &= p_1r_1 \Gamma_2(t) + p_2r_2 \Gamma_1(t)  - (p_1+p_2)\delta(t),
\end{align}
\end{subequations}
and
\begin{subequations}
\begin{align}
\alpha(t) &\equiv \sum_n n \Big(  \bra{01,n} \rho \ket{10,n+1} - \bra{10,n+1} \rho \ket{01,n} \Big) \\
\beta(t) &\equiv \sum_n n \Big(  \bra{00,n} \rho \ket{00,n} + \bra{01,n} \rho \ket{01,n} \Big) \\
\gamma(t) &\equiv \sum_n (n-1) \Big(  \bra{00,n} \rho \ket{00,n} + \bra{10,n} \rho \ket{10,n} \Big) \\
\delta(t) &\equiv \sum_n   \bra{00,n} \rho \ket{00,n}.
\end{align}
\end{subequations}

\begin{figure*}
  \centering
       {\LP{Extractable work fluctuations and  extractable} work relative to fluctuations        }

 \vspace{21pt}
        \includegraphics[trim=00 00 00 00,width=0.37\textwidth]{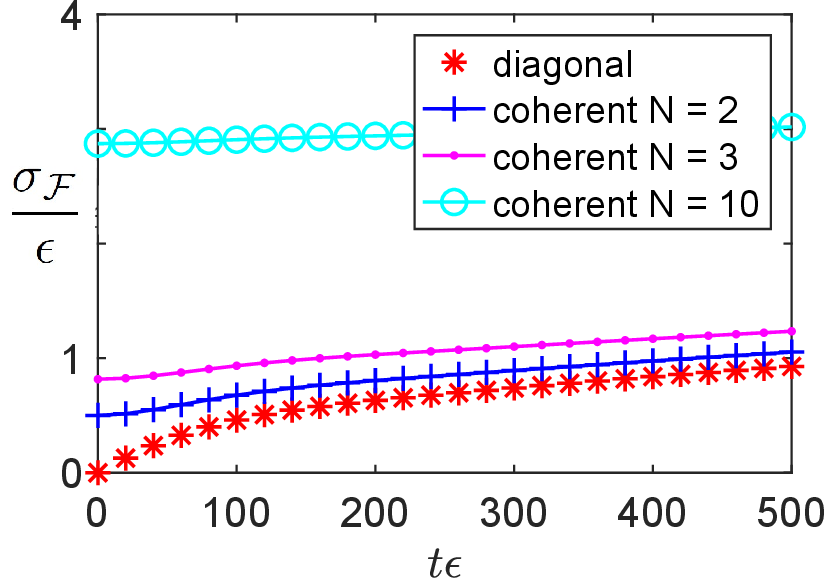} 
     \hspace{26pt}   \includegraphics[trim=00 00 00 00,width=0.35\textwidth]{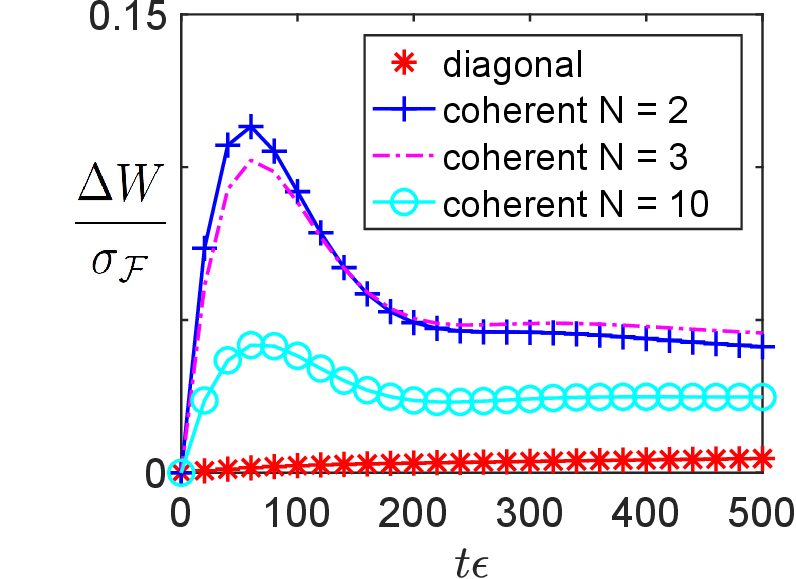}         
        
      \caption{\label{fig:workfluctuations}
\LP{Extractable work fluctuations $\sigma_\work$ (left) and extractable} work relative to the fluctuations $\Delta W / \sigma_\work$ (right) as a function of time for different initial states, with $p_1 = p_2 = g/\epsilon$, $E_1/T_c = 0.51$,  $E_2/T_h = 0.50$, and $\theta = \pi/2$. 
The fluctuations in the \LP{extractable work} for the incoherent state $\rho_\textnormal{diag}$ (red asterisk) are zero initially, but grow in time. For a coherent superposition between few (blue cross and purple dot) eigenvalues of $H_\bat$ the initial fluctuations are significantly higher than for $\rho_\textnormal{diag}$, but the two become comparable for longer times. Since the engine produces a much higher \LP{extractable} work output for the coherent state (see Fig.~\ref{fig-app:powerandwork}), the \LP{extractable}  work relative to the fluctuations is considerably higher for the coherent state.
Taking coherent superpositions between many states (lightblue circle) gives  a further increase the power fluctuations, which results in a lower \LP{extractable}  work relative to the fluctuations. 
           }
\end{figure*}

Figure~\ref{fig:workfluctuations} depicts the fluctuations of the \LP{extractable work} as a function of time for $\rho_\textnormal{diag}$ and $\ket{\Psi_N}\!\bra{\Psi_N}$ as initial states, for different values of $N$. As expected, coherent initial states, which have higher initial fluctuations, lead to final states with higher fluctuations than the diagonal state.
However, the \LP{extractable} work increases with coherent superpositions as well, resulting in \LP{an extractable} work relative to the fluctuations that is considerably higher for coherent superpositions between few states ($N = 2$ and $N = 3$) than for the incoherent state. Further increasing the number of states in the superposition leads to a decrease in the \LP{extractable} work relative to the fluctuations (although still higher than for $\rho_\textnormal{diag}$). Therefore, states in coherent superposition between few levels seem the most desirable: the \LP{extractable} work greatly increases, while keeping relative fluctuations low.

This raises the discussion of the `quality' of the \LP{extractable} work.
If one takes the average energy of the battery as the sole figure of merit for the work storage protocol, then considering states in coherent superpositions of energy eigenstates is better than diagonal states, resulting in higher charging power.
On the other hand, for superpositions between many states the fluctuations of the \LP{extractable} work increase considerably.  
On the other hand, when the fluctuations of the \LP{extractable} work are taken into account, taking superpositions between few levels is better, resulting in an increase in the final amount of \LP{extractable} work without considerably higher fluctuations than for a diagonal initial state.

\end{document}